\documentclass[12pt,preprint]{emulateapj}

\shorttitle{H$_2$ Emission in the Taffy Galaxies}
\shortauthors{Peterson et al.}

\begin{document}

\newcommand{\Hi}{\ion{H}{1} }
\newcommand{\Hii}{\ion{H}{2} }
\newcommand{\Ht}{${\rm H_{2}}$ }
\newcommand{\mic}{$\micron$ }

\title{Detection of Powerful Mid-IR H$_2$ Emission in the Bridge Between the Taffy Galaxies}

\author{B. W. Peterson\altaffilmark{1,2}, P. N. Appleton\altaffilmark{3}, G. Helou\altaffilmark{2}, P. Guillard\altaffilmark{4}, T. H. Jarrett\altaffilmark{2}, M. E. Cluver\altaffilmark{4,5}, P. Ogle\altaffilmark{4}, C. Struck\altaffilmark{1}, F. Boulanger\altaffilmark{6}}

\altaffiltext{1}{Department of Physics and Astronomy, Iowa State University, Ames, IA 50011}
\altaffiltext{2}{IPAC, California Institute of Technology, Pasadena, CA 91125}
\altaffiltext{3}{NASA Herschel Science Center, IPAC, California Institute of Technology, Pasadena, CA 91125}
\altaffiltext{4}{Spitzer Science Center, IPAC, California Institute of Technology, Pasadena, CA 91125}
\altaffiltext{5}{Australian Astronomical Observatory, P.O. Box 296, Epping, NSW, 1710, Australia}
\altaffiltext{6}{Institut d'Astrophysique Spatiale, Universit\'{e} Paris Sud 11, Orsay, France}

\begin{abstract}

We report the detection of strong, resolved emission from warm \Ht in the Taffy galaxies and bridge. Relative to the continuum and faint PAH emission, the \Ht emission is the strongest in the connecting bridge, approaching $L$({H$_2$})/$L$(PAH8$\micron$) = 0.1 between the two galaxies, where the purely rotational lines of \Ht dominate the mid-infrared spectrum in a way very reminiscent of the group-wide shock in the interacting group Stephan's Quintet. The surface brightness in the 0--0 S(0) and S(1) \Ht lines in the bridge is more than twice that observed at the center of the Stephan's Quintet shock. We observe a warm \Ht mass of $4.2\times10^{8}$~$M_\odot$ in the bridge, but taking into account the unobserved bridge area, the total warm mass is likely to be twice this value. We use excitation diagrams to characterize the warm molecular gas, finding an average surface mass of $\sim$$5\times10^6$~$M_{\odot}$~kpc$^{-2}$ and typical excitation temperatures of 150--175~K. \Ht emission is also seen in the galaxy disks, although there the emission is more consistent with normal star forming galaxies. We investigate several possible heating mechanisms for the bridge gas, but favor the conversion of kinetic energy from the head-on collision via turbulence and shocks as the main heating source. Since the cooling time for the warm \Ht is short ($\sim5000$ yr), shocks must be permeating the molecular gas in bridge region in order to continue heating the ${\rm H_2}$.

\end{abstract}

\keywords{galaxies: individual (UGC 12914, UGC 12915) --- galaxies: interactions --- intergalactic medium}

\section{Introduction}

The \emph{Spitzer Space Telescope} has led to tremendous advances in the study of galaxies and their interstellar media. One particularly fruitful area has been the study of the lower-level purely rotational emission lines of molecular hydrogen. An interesting example was the detection by \citet{app06} of powerful emission from purely rotational \Ht lines in the group-wide shock of the strongly interacting group Stephan's Quintet (SQ). Spectral mapping reveals that the emission is distributed along the whole shock and exceeds the X-ray power by a factor $\geqslant 3$, indicating that \Ht line emission is an important cooling process in the post-shock region \citep{clu10}.

Powerful \Ht emission has also been detected in many other environments, including luminous infrared galaxies
\citep{lutz03}, radio galaxies \citep{ogl07, ogl10}, AGNs \citep{rou07}, and cool cluster core galaxies \citep{ega06, don11}. In
another case \Ht is detected in the apparent wake of a bow-shock in the cluster Abell 3627 \citep{siv09}. The class of strong molecular hydrogen emission galaxies (called MOHEGs; \citealt{ogl07}) is defined by the strength of the \Ht 0--0 S(0)--S(3) lines relative to that of polycyclic aromatic hydrocarbons (PAHs), with MOHEGs having $L({\rm   H_2})/L$(PAH8$\micron$) $> 0.04$ \citep{ogl10}. The mechanism that powers the \Ht emission in MOHEGS is not fully understood, in part because the emission regions are usually unresolved, but in the radio galaxies is likely to be jet-driven shocks \citep{ogl10, nes10}. The SQ shock is exceptional because it is spatially extended over $\sim$30~kpc and associated with a well-studied intergalactic shock, which has led to a clear picture of the heating mechanism as dissipation of the kinetic energy associated with the collision \citep{gui09}.

Here we report the detection of another system containing strong, resolved ${\rm H_2}$ emission in the bridge between the interacting galaxies UGC 12914/5, also known as the ``Taffy'' galaxies because of the extended radio emission stretched between the two disks \citep{con93}. The emission closely resembles that observed in SQ, with the H$_2$ lines featuring prominently in the mid-IR spectrum. 

The Taffy system is believed to be the result of a nearly head-on collision between two disk galaxies. The galaxies, UGC 12914 and 12915, have heliocentric velocities of $4371\pm8$ and $4336\pm7$~km~s$^{-1}$ respectively, and we adopt a distance of 60~Mpc based on a Hubble constant of 72~km~s$^{-1}$~Mpc$^{-1}$. The disks are $\sim$12~kpc apart and are believed to be separating mainly in the plane of the sky at a velocity  of 450~km~s$^{-1}$ \citep{con93}. This estimate is based on assumed masses and a parabolic orbit, and may be an upper limit since it does not take into account dynamical friction resulting from the overlapping dark matter halos. Among the notable features of the system is the gas-rich bridge, which contains about 25\% of the \ion{H}{1} gas in the system \citep{con93}. The bridge is also host to a significant amount of molecular gas \citep{smi01, bra03, gao03} and cold dust \citep{zhu07}. A warm dust component is apparently heated by UV radiation from the disks \citep{jar99}. H$\alpha$ images reveal that the bridge has little star formation outside of a single large \ion{H}{2} region near UGC 12915 \citep{bus90}. 

When two galaxy disks collide nearly face-on at high velocity, we expect widespread collisions between diffuse atomic clouds, which have a large covering factor in galaxy disks \citep{ler08, san10}. In the collision, dense molecular clouds in one galaxy will ram through the more diffuse atomic gas of the second galaxy. The coupling of the generally smaller molecular clouds to larger areas via magnetic fields will increase the drag on the galaxies, enhancing the interaction. The net result of these processes is that the atomic gas will be splashed out into the bridge with a range of transverse velocities with values extending up to the relative collision velocity \citep{str97}. Some of this gas will fall promptly back into both disks, and some will be stretched between them. Large molecular clouds will tend to stay close to their parent disks, which partially explains the lack of star formation in the bridge. The gas splashed into the bridge will also have a broad range of angular momenta, resulting in continuing cloud collisions that produce shocks and turbulence.

This paper is primarily aimed at understanding the distribution and excitation of warm molecular hydrogen in the Taffy galaxies and bridge, and is organized as follows. In Section~\ref{sec:obs} we describe the observations and data reduction. In Section~\ref{sec:results} we present the mid-IR spectra, estimate the cooling rate, and show how the \Ht emission is distributed in the Taffy system. We also determine the properties of the warm ${\rm H_2}$. In Section~\ref{sec:discussion} we discuss possible heating mechanisms for the warm ${\rm H_2}$, and compare with the \Ht mass inferred from previous CO observations. We briefly summarize our findings in Section~\ref{sec:summary}.

\section{Observations and Data Reduction} \label{sec:obs}
\subsection{IRS spectra}

\begin{figure}
\epsscale{.70}
\plotone{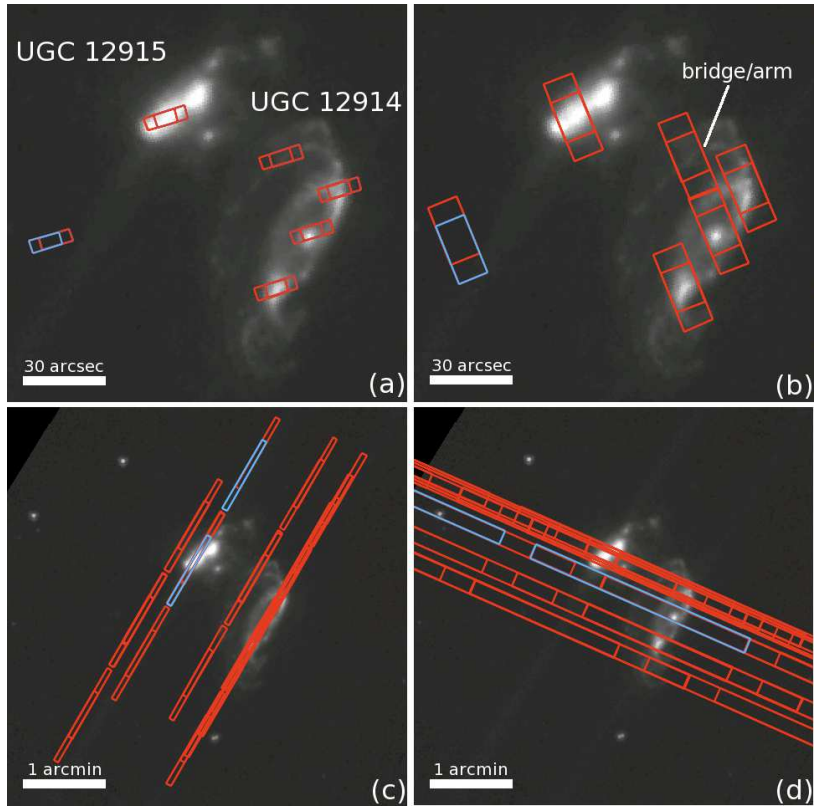}
\caption{IRS SH (a), LH (b), SL (c), and LL (d) slit positions overlaid on IRAC 8.0~$\micron$ image. Each module was nodded between two positions at each target, with the target at the 1/3 and 2/3 positions in the slit. A single nod position for each module is shown in blue for clarity. In (b), we also label the position we call the ``bridge/arm'' in the text. For the SL and LL modules, both orders are shown. North is up and east to the left.}\label{fig:aor}
\end{figure}

We obtained spectra of the UGC 12914/5 system from the \emph{Spitzer} public archives (Program ID: 21, PI: J. Houck). The observations were made with the Infrared Spectrograph (IRS; \citealt{hou04}) on board the \emph{Spitzer Space Telescope} \citep{wer04} on 2005 July 8 and 10 using the Short-Low (SL; 5.2--14.5~$\micron$), Long-Low (LL; 14.0--28.0~$\micron$), Short-High (SH; 9.9--19.6~$\micron$), and Long-High (LH; 18.7--37.2~$\micron$) modules. The positions of the slits for all four modules are shown in Figure~\ref{fig:aor}. The AORs were designed to target the nuclei and selected regions of the galaxies, not specifically the bridge. However, we were able to exploit the partially overlapping nature of the low resolution slits to make sparse maps of the region.

The observations were initially processed by \emph{Spitzer} Science Center (SSC) pipeline version S18.7.0. The starting point of our analysis were the Basic Calibrated Data (BCD) frames from the science pipeline. The BCD frames are a high-level calibrated data product which represent slope-fitted data values in electron~s$^{-1}$ derived from corrected data. Corrections include saturation, non-linearity, detector droop and stray-light corrections. BCD frames associated with each target position were coadded and background subtracted using dedicated off observations. In the SL and LL modules the dedicated off observations were combined with spectral orders far from the galaxies. Rogue pixels, especially common in the LH module due to cosmic-ray activation of pixels, were identified manually by blinking the two nod positions against each other to help distinguish rogue pixels from real data. The rogue pixels were then replaced by suitable averaged values from adjacent pixels using customized software.

\begin{figure}
\epsscale{.50}
\plotone{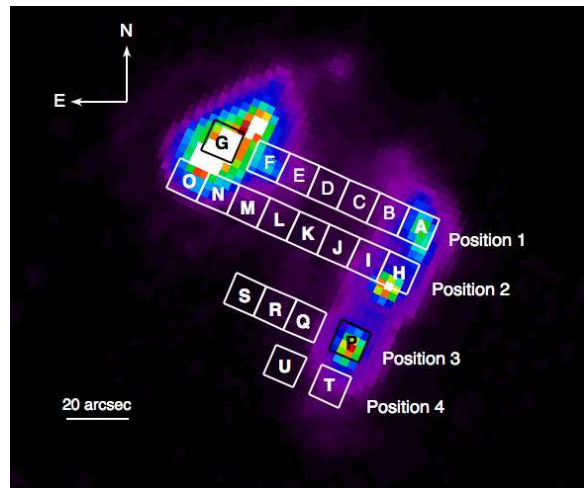}
\caption{Spectral extraction regions A--U overlaid on MIPS 24~$\micron$. The extraction regions are $10\farcs15\times10\farcs15$. Each row of extraction regions corresponds to one of the LL slit positions, which are labeled.} \label{fig:cubism}
\end{figure}

Spectra were obtained at two nod positions for each target, with the target placed at the 1/3 and 2/3 positions in the slit. The CUBISM software \citep{smi07a} was used to construct partial spectral maps along the LL slits, allowing extractions from numerous positions within the system, labeled A--U as shown in Figure~\ref{fig:cubism}. The extraction regions had angular size $10\farcs15\times10\farcs15$. The SL slits are oriented orthogonally and partially overlap in regions C and J. In both cases, the SL spectra covered about 1/3 of the area of the LL, and were scaled up to compensate for this difference. This rescaling assumes that the emission is uniform over the aperture, which introduces uncertainty into measurements of the SL lines in these regions. We therefore use these lines only as a check on our excitation diagrams (see Section~\ref{sec:exc}). Spectra from the LL and SL slits will be referred to herafter as low-res spectra (resolving power $R$ between 57--127). 

The high-resolution spectra (hereafter hi-res with $R = 600$) were extracted with the SSC software SPICE, using the point source calibration for the galaxies and extended source calibration for the bridge. The choice of calibrations involves some uncertainty, since the targets are neither fully resolved point sources nor perfectly uniform extended sources. In the worst case, the difference between a point source and extended source extraction (a wavelength-dependent slit-loss correction factor) can lead to $\sim$40\% difference in extracted line fluxes for the lines observed. 

Line fluxes in each extraction region were measured using the SMART software package \citep{hig04}. In regions C and J, the partial overlap with the SL module expanded the wavelength coverage. In these cases we measured the line fluxes using both SMART and PAHFIT \citep{smi07b}. PAHFIT fits known lines, bands, and dust contiua to a stitched version of the low-res spectra. Similar line fluxes were obtained with both methods; results are presented in Table~\ref{tab:h2}. We note a slight positive offset in the measured line fluxes in region J by PAHFIT relative to SMART, but the differences are only at the 2$\sigma$ level. The effect is more pronounced in the 28~$\micron$ line. As we point out in the footnote to Table~\ref{tab:h2}, the LL1 spectrum from which this was derived showed some inconsistency in its continuum level compared with LL2, so this line is uncertain no matter how it is measured.

The hi-res line fluxes were measured by single Gaussian fitting using SMART. The LH slit covers an area $22\farcs3\times11\farcs1$. The SH slit covers a smaller area of  $11\farcs3\times4\farcs7$. The SH spectra were scaled at each nod position so that the continuum matched between the two modules. The scaling factors were similar for both nods. The line fluxes and mean of the scaling factors for the two nods are presented in Table~\ref{tab:h2}.

\subsection{IRAC and MIPS images} \label{sec:phot}

Images of the system obtained with the Infrared Array Camera (IRAC; \citealt{faz04}) and Multiband Imaging Photometer for \emph{Spitzer} (MIPS; \citealt{rie04}) were available from the \emph{Spitzer} public archives (Program ID: 21, PI: J. Houck). These data were processed by SSC pipeline version S18.7.0, and were of sufficient quality that further processing was not required.

Photometric measurements were made in each of the LL bridge extraction regions using square apertures with the IRAF task polyphot. The CUBISM spectral extractions use extended source calibration, so the photometric measurements were not aperture-corrected. The IRAC 3.6, 4.5, 5.8, and 8.0~$\micron$ images were Gaussian-convolved to match the resolution of the MIPS 24~$\micron$ image, which is comparable to the resolution at the \Ht S(0) 28.22~$\micron$ line.  For comparison with the hi-res spectra, we also measured the images using a set of boxes matched to each of the LH nod positions.

\section{Results} \label{sec:results}
\subsection{Spectra}

\begin{figure}
%\epsscale{.50}
\plotone{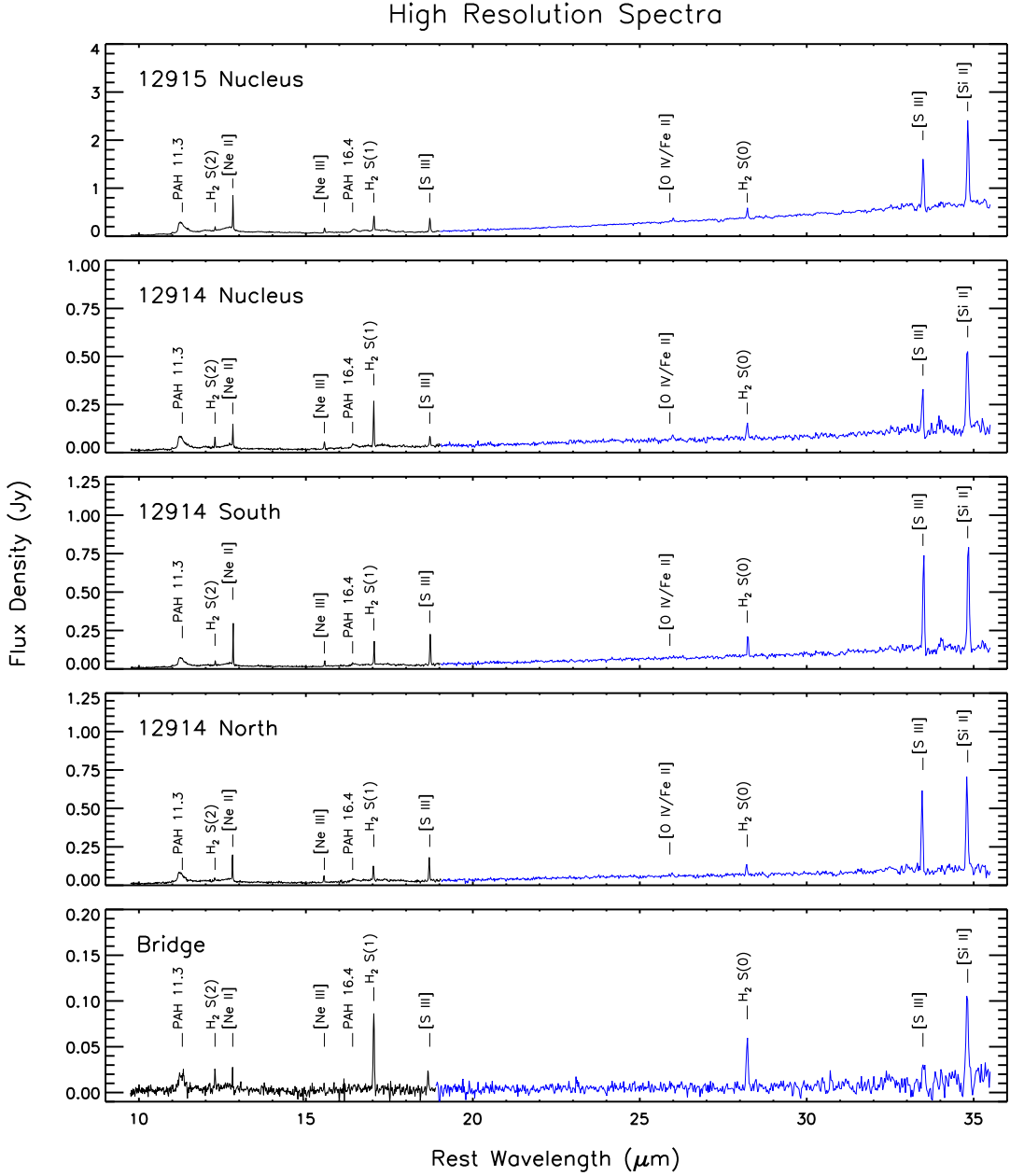}
\caption{High resolution spectra from the IRS SH (black) and LH (blue) modules. The SH spectra were scaled up to match the LH continua. The mean scaling factors for the two nods, along with the measured line fluxes, are presented in Table~\ref{tab:h2}.}\label{fig:hi}
\end{figure}

The hi-res spectra of the nuclei and the bridge are shown in Figure~\ref{fig:hi}. The strength of the \Ht lines is very striking, especially in the bridge, where the 0--0 S(0) and S(1) lines are stronger than all other lines except the [\ion{Si}{2}]34.82~$\micron$ line. The S(1) line is similarly powerful in the UGC 12914 nucleus. The \Ht line fluxes are presented at the top of Table~\ref{tab:h2}, while the fine structure line fluxes are shown in Table~\ref{tab:fine}. We will further quantify the strength of the \Ht
emission in Section~\ref{sec:heating}.

\begin{figure}
%\epsscale{.50}
\plotone{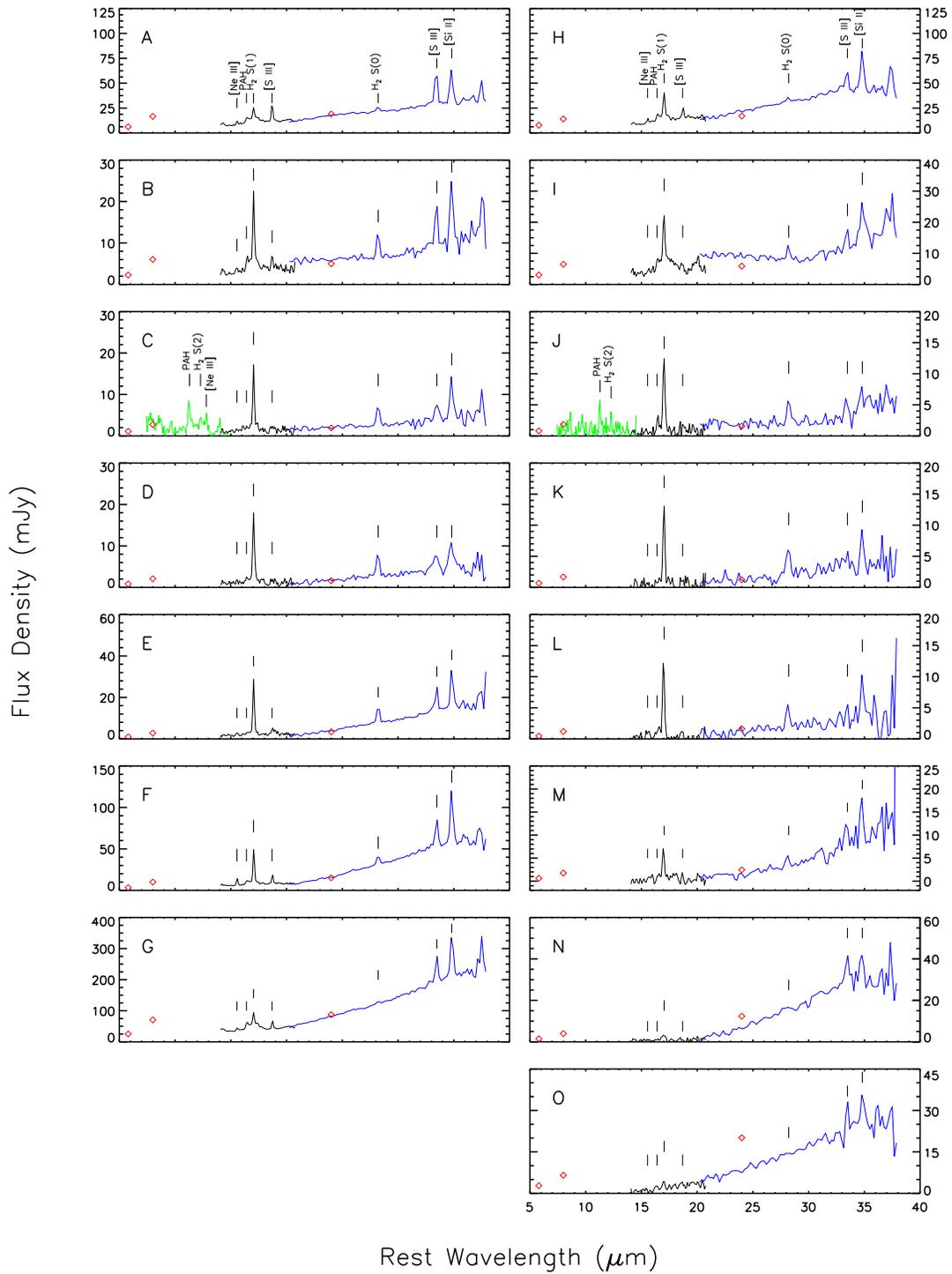}
\caption{Low resolution spectra from IRS LL1 (blue), LL2 (black), SL1 (green, where available) modules extracted from bridge regions A--G (position 1; left column) and H--O (position 2; right column). The SL1 data have been scaled up to the same aperture size as the LL1 assuming a uniform distribution over the aperture. Photometric data points from IRAC and MIPS are shown as red diamonds.}\label{fig:lo1}
\end{figure}

\begin{figure}
%\epsscale{.50}
\plotone{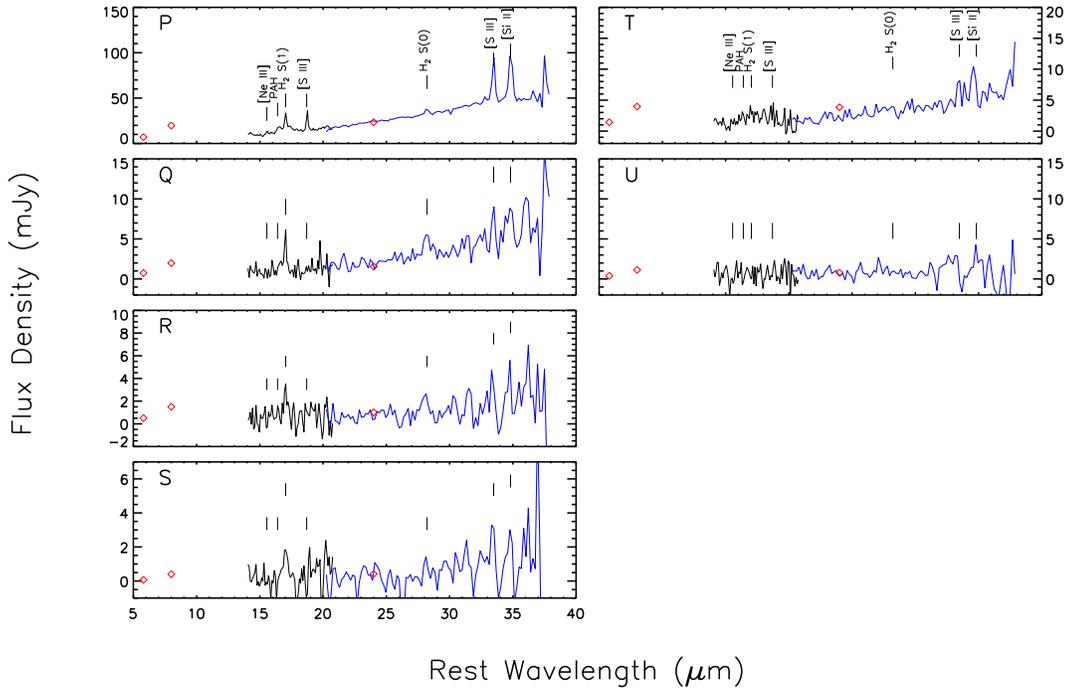}
\caption{Low resolution spectra from IRS low resolution extraction regions P--S (position 3; left column) and T--U (position 4; right column). Symbols are as in Figure~\ref{fig:lo1}.}\label{fig:lo2}
\end{figure}

The low-res spectra show that the emission in the bridge is not confined to the position of the high resolution bridge/arm slit which, as shown in Figure~\ref{fig:aor}, also includes the UGC 12914 spiral arm. The \Ht S(1) line is measured in regions A--S. The two regions where it could not be measured (T and U) are near the southern end of UGC 12914 and lie outside the apparent extent of the bridge, based on the radio contours of \citet{con93}. Spectra from all of the regions are shown in Figures~\ref{fig:lo1} and \ref{fig:lo2}. The red diamonds show the continuum flux densities measured from IRAC and MIPS for the same extraction regions. Note that in almost all cases these broadband measurements show good agreement with the spectra, indicating that the relative scaling made between spectral orders was appropriate. In only one case (region O) do the IRAC fluxes disagree with the extracted spectra. However, since no lines were detected in that position, the discrepancy does not effect our analysis of the ${\rm H_2}$.

\subsection{\Ht distribution}

\begin{figure}
%\epsscale{.50}
\plotone{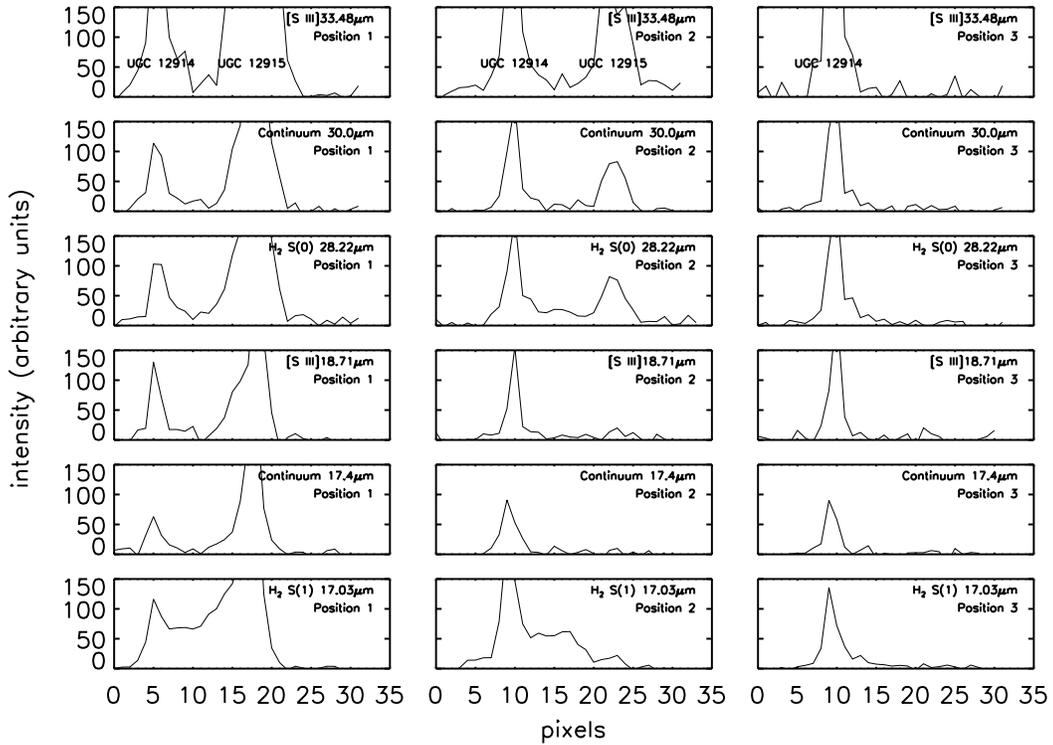}
\caption{Cross sections through the LL1 and LL2 slits at the wavelengths of [\ion{S}{3}]33.48$\micron$, 30~$\micron$ continuum, \Ht S(0) 28.22~$\micron$, [\ion{S}{3}]18.71$\micron$, 17.4~$\micron$ continuum, and \Ht S(1)~17.03~$\micron$. The two continuum profiles give a sense of the the galaxy widths at wavelengths near the lines of interest. The positions are indicated in Figure~\ref{fig:cubism}. In position 2, the slit is slightly south of UGC 12915. These plots, particularly of the \Ht S(1) line, demonstrate that \Ht emission is extended across the full width of the Taffy bridge.}
\label{fig:xSec}
\end{figure}

We now present two alternative ways of demonstrating that the \Ht emission, as measured in the low-res spectra, extends between the galaxies in the bridge. Figure~\ref{fig:xSec} shows cross sections of the LL1 and LL2 slits which cut through the system along three slices, which we call positions 1--3. The locations of the galaxies at each position are shown in the top panels. No significant emission was detected at position 4, so it is not shown. 

Several important results emerge from the cross-sectional analysis. First, both the  [\ion{S}{3}]33.48$\micron$ and the continuum emission (measured at wavelengths of 30~$\micron$ and 17.4~$\micron$) are concentrated in the galaxies, with little originating in the bridge. This contrasts with the \Ht S(0) and S(1) lines, which show significant emission between the galaxies, especially at position 2. The \Ht S(1) line is even stronger in the bridge compared to the galaxies, and actually drops significantly just south of UGC 12915. The right panels of Figure~\ref{fig:xSec} show that the emission is much weaker outside of UGC 12914 at position 3 than at position 2 (see also Figures~\ref{fig:lo1} and \ref{fig:lo2}), indicating that either the gas density or temperature drops significantly south of a line connecting the nuclei. 

\begin{figure}
\epsscale{.85}
\plotone{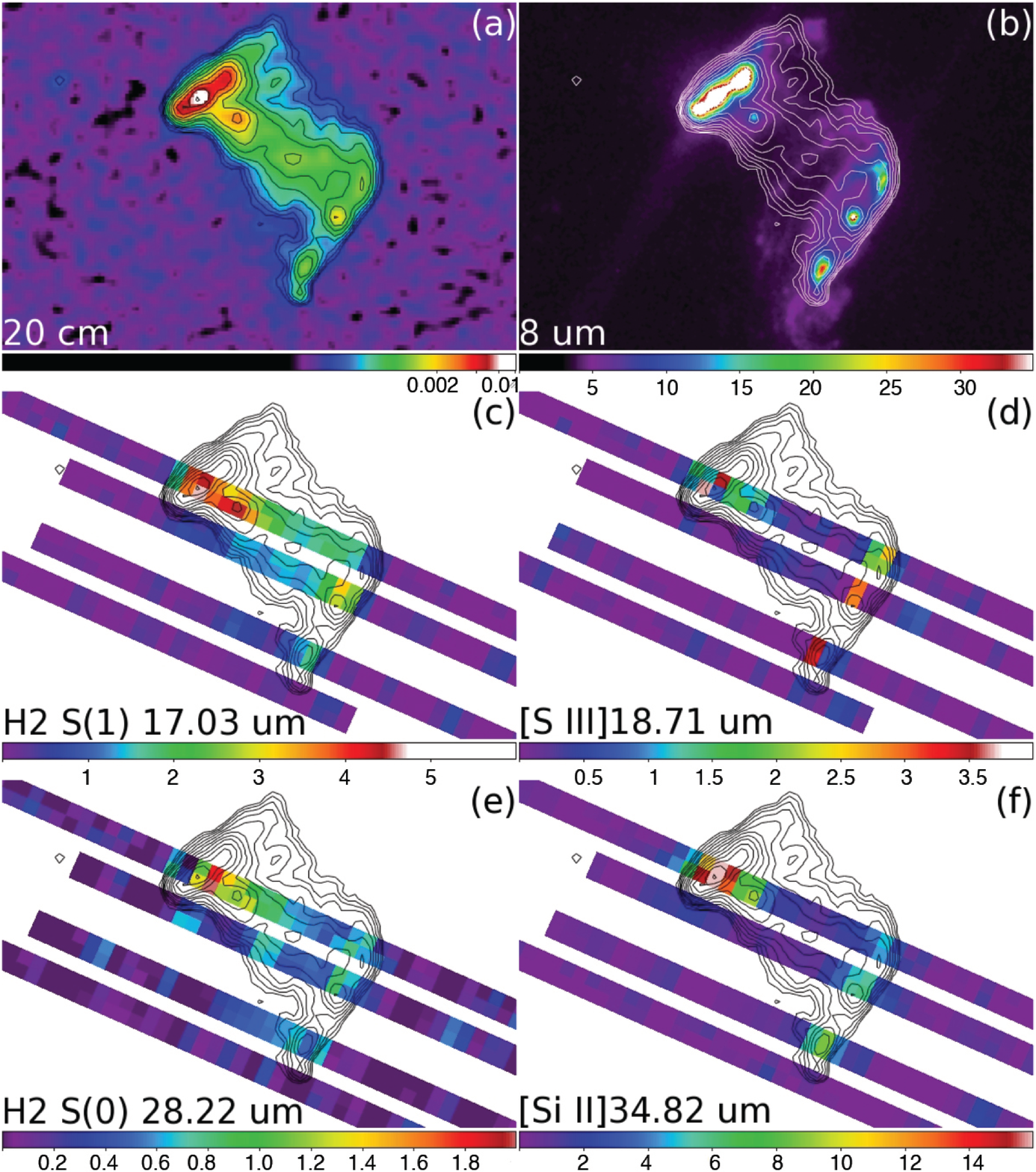}
\caption{(a) 20 cm radio image \citet{con93}, with logarithmic color scale and contours from 0.1--7~mJy~beam$^{-1}$. (b) 20 cm contours are also shown overlaid on IRAC 8.0~$\micron$ image (b), sparse maps of \Ht S(1) 17.03~$\micron$ line (c), [\ion{S}{3}18.71]$\micron$ line (d), \Ht S(0) 28.22~$\micron$ line (e), and [\ion{Si}{2}]34.82~$\micron$ line (f). The sparse maps of the S(0) and S(1) lines show that the \Ht emission is extended across the entire bridge, compared to the emission from [\ion{S}{3}] and [\ion{Si}{2}], which have comparable wavelengths but do not show emission outside the galaxies. The units on plots (b)--(f) are MJy~sr$^{-1}$.}\label{fig:maps}
\end{figure}

The extended nature of the \Ht emission in the bridge can also be demonstrated using sparse spectral maps. Although the Taffy system was not fully mapped, sufficient long-slit low-res spectra were obtained to allow the construction of sparse LL spectral maps using CUBISM, which are shown in Figure~\ref{fig:maps}. To provide the reader with a better visualization of the results, the S(0) and S(1) data (color image) are overlaid with 20~cm radio continuum (contours) from \citet{con93}. We also show these radio contours overlaid on an IRAC 8.0~$\micron$ image, and sparse maps of the [\ion{S}{3}]18.71$\micron$ and [\ion{Si}{2}]34.82$\micron$ lines. These latter lines are selected because they have wavelengths comparable to the \Ht S(1) and S(0) lines, and so provide a sense of the contamination level in the bridge as a result of emission from the galaxies. Figure~\ref{fig:maps} shows that extended \Ht emission is present in the bridge region, similar to the distribution of dust \citep{jar99, zhu07} and CO \citep{gao03}, though differing from the \ion{H}{1}, which peaks in the bridge \citep{con93}.

\subsection{Total \Ht cooling rate} \label{sec:cool}

The total \Ht luminosity (or equivalently, the \Ht line cooling rate) in various regions of the bridge, and an estimate for the whole bridge, can be calculated from the \Ht line fluxes. Since the spatial coverage is more complete for the low-res spectra, we restrict our discussion to those data, which cover the S(0) and S(1) lines only, with the exception of regions C and J. 

To determine the total 0--0 S(0) and S(1) line luminosity in the bridge, we exclude the regions obviously on the disk of the galaxies (A, G, H, P, and T), and region F (which contains the star forming knot), as well as regions very near the galaxies, where contamination from the disk is likely to be largest (B and I). This leaves regions C, D, E, J, K, L, M, Q, and R as bridge spectra. The total flux from these regions is $5.8\times10^{-16}$~W~m$^{-2}$, giving a total luminosity of 1.16$\times10^{34}$~W over a projected area of 78.5~kpc$^2$. The mean surface brightness is $\sim$$1.5\times10^{32}$~W~kpc$^{-2}$.\footnote{For the hi-res region that lies in the bridge (called ``bridge/arm'' in Table 1) we note that the surface brightness of the H2 0-0 S(0) and S(1) lines, averaged over the two LH nod positions (covering an area of  $29\farcs7 \times 11\farcs1$) is $1.0\times10^{32}$~W~kpc$^{-2}$. This is lower than the surface brightness seen in region C measured in the low-res spectra. Although we do not have a complete explanation for the discrepancy, the result may indicate that the emission is clumpy on the scale of the LH slit. We note that surface brightness variations comparable to this difference are seen  on the scale of the low-res apertures ($\sim$$10\arcsec$), which may partly explain the difference.} 

To estimate the total \Ht luminosity, we must also estimate the spatial extent of the emission. We assume that the \Ht emission is roughly bounded by the extent of the radio emitting bridge. We define this area as being bounded by the extent of the radio continuum (down to a level of 0.1~mJy in the \citealt{con93} 20~cm emission map) assuming reasonable inner edges to the galaxy disks. This provides a bridge area of 170~kpc$^2$, and gives a total bridge \Ht luminosity $L({\rm H_2}) = 2.6(\pm0.7) \times10^{34}$~W, assuming the properties of the unobserved bridge are similar to those of the observed component. In regions C and J where the SL data allows the addition of the S(2) and S(3) lines, the extra contributions amount to an increase of 28\% in the luminosity. Our estimate of the total warm \Ht luminosity is thus likely to be a lower limit.

\subsection{Excitation diagram and \Ht mass surface densities}  \label{sec:exc}

Although we do not have full spectral coverage of the whole bridge, we can explore the variation of \Ht properties as a function of position in the bridge where we have useful data. The extraction regions A--U, defined in Figure~\ref{fig:cubism}, are generally restricted to the LL module of the IRS, and thus cover the only 0--0 S(0) and S(1) lines. As previously discussed, regions C and J also provide information for the S(2) and S(3) lines at 12.3 and 9.66~$\micron$. These measurements are sufficient to allow us to make a preliminary exploration of the excitation properties of the warm \Ht across the bridge, subject to several constraints and assumptions discussed below.

We have constructed \Ht excitation diagrams for the regions A--U. These diagrams plot the column density $N_{\rm u}$ of \Ht in the upper level of each transition, normalized by its statistical weight, versus the upper level energy $E_{\rm u}$ \citep[e.g.,][]{rig02}, which were derived from the measured fluxes assuming local thermodynamic equilibrium (LTE) for each position. Since most of the excitation diagrams consist of only two points, corresponding to the 0--0 S(0) and S(1) lines, we do not show them here, but rather present in Table~\ref{tab:excipars} the results for a single-temperature fit through the two points in each region. These fits provide a baseline measurement, and allow us to explore how the ratio of the S(0) and S(1) lines might vary as a function of position in the system. For regions N, O, and S only an upper limit was obtained for the S(0) line, so no temperature was derived. No H$_2$ lines were detected in region U.

The results of this analysis are presented in Table~\ref{tab:excipars}. Most of the regions in the bridge have temperatures (represented by the slope in the excitation diagram) in the range 157--175~K, with only two regions (Q and R) having lower temperatures of 130--133~K. The nuclei of both galaxies (regions G and H) show a higher temperature. The UGC 12914 nucleus has a \Ht temperature of 195$\pm$12~K, some 30~K warmer than the average bridge region. The warm \Ht surface density derived from these measurements (also given in Table~\ref{tab:excipars}) range from $10^7$~$M_{\odot}$~kpc$^{-2}$ at position F, which contains a star forming knot, to the lowest value of $2.4\times10^6$~$M_{\odot}$~kpc$^{-2}$ at region M. The mid-bridge regions D, K, and R have similar \Ht surface densities of 5.0--6.6$\times10^6$~$M_{\odot}$~kpc$^{-2}$. We again select regions C, D, E, J, K, L, M, Q, and R as representing the bridge. The mean and median value of the surface density over regions between the galaxies is $5.3\pm0.5$ and $5.5\times10^6$~$M_{\odot}$~kpc$^{-2}$, respectively. Based on the cooling rate from the same regions (see Section~\ref{sec:cool}) and an assumed average temperature of 166~K, the average thermal cooling time for this warm mass density is $\sim$5000~yr, which is very short compared with the bridge formation time.

Summing the observed warm \Ht masses over these representative bridge regions (78.5~kpc$^2$ in area) yields an observed mass of $4.2\times10^{8}$~$M_{\odot}$. We estimate that the total warm \Ht mass for the whole bridge area of $\sim170$~kpc$^2$ could be as large as 9(+2/$-5$)$\times10^8$~$M_{\odot}$ if the entire bridge has similar properties to the observed bridge regions. The lower limit comes from the extreme assumption that no new \Ht is detected in the unobserved area.

We have made several simplifying assumptions in estimating the \Ht surface density from these data. First, we have assumed a thermalized equilibrium value for the ortho-to-para ratio. For temperatures above 300~K, this ratio is 3, but for temperatures in the range 130--180~K this value varies from 2.3--2.8 respectively (see Equation 4 of \citealt{wil00}). Under most circumstances we might expect the lower-$J$ transitions to be in equilibrium. Normally, the ortho-to-para ratio would be investigated by looking for systematic differences in the odd and even transitions in the excitation diagram. However, because we have so few points (in most cases only two) we must simply assume equilibrium values. This obviously introduces an unknown uncertainty in the final derived properties of the gas. Deviations from LTE would lead to uncertainties in the assumed excitation temperature of the gas and thus the final total column densities. In regions C and J where four lines are detected (discussed below) we do not see any obvious deviations in the odd and even values for $N/g$ versus upper-level energy, but this trend may not hold in all bridge regions. 
%In Stephan's Quintet, where many more transitions were detected, and which seems to have strong similarities with our current data, we find no obvious deviations from LTE \citep{clu10}.

\begin{figure}
\epsscale{.70}
\plotone{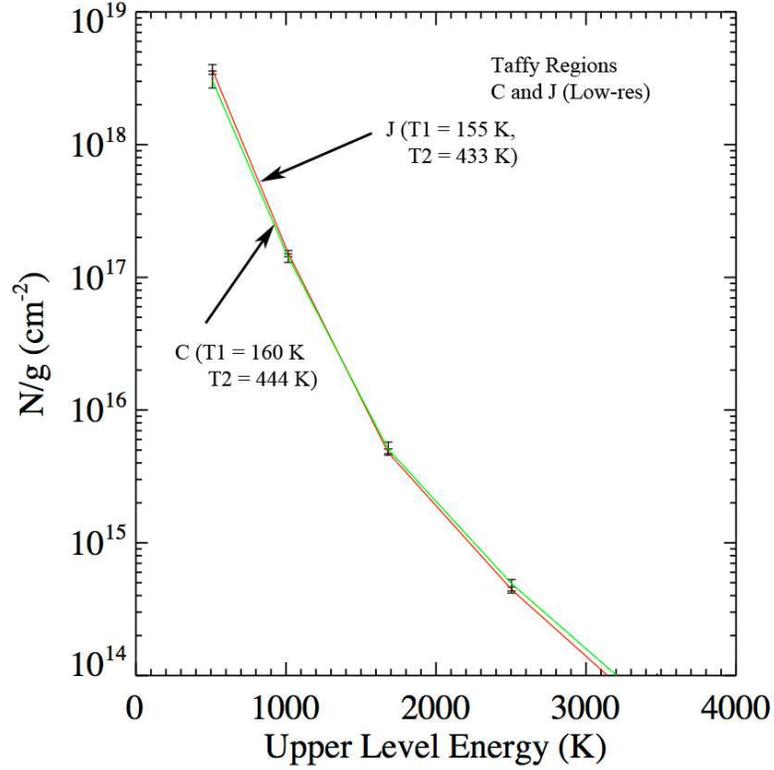}
\caption{Excitation diagrams with multi-temperature fits of low-res regions C (green) and J (red). The fit parameters are shown in Table~\ref{tab:excipars}.}\label{fig:excite2}
\end{figure}

A second assumption is that the 0--0 S(0) and S(1) transitions can be fitted by a single temperature component. Excitation diagrams for \Ht observations are usually fitted by several temperature components \citep[e.g.,][]{rou07}. To explore this assumption we use the low-res regions C and J, where we were able to measure the S(2) and S(3) lines, to investigate how this assumption affects the results. The fits for these regions are shown in Figure~\ref{fig:excite2}, while the gas parameters are presented in Table~\ref{tab:excipars}. We find that in addition to a cool component, a warmer component ($\sim$430--440~K) is needed to explain the S(2) and S(3) measurements. However, the effect on the temperature of the coolest component, which dominates the mass surface density, is quite small. 

\begin{figure}
\epsscale{.70}
\plotone{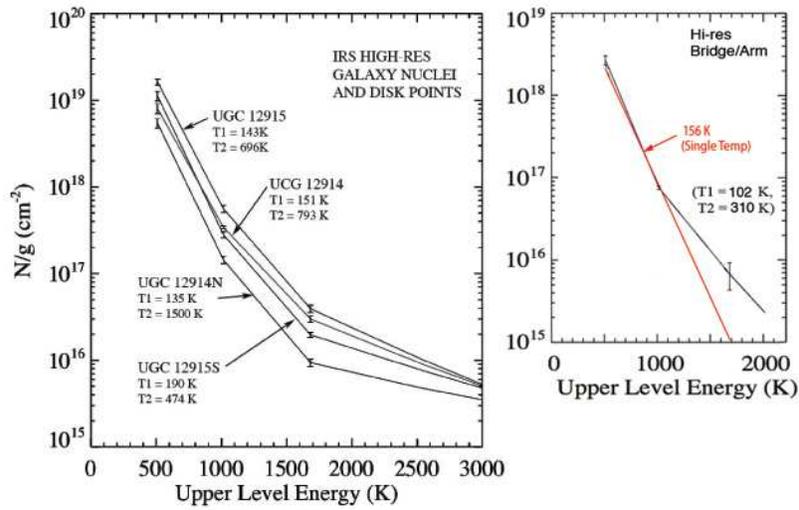}
\caption{Excitation diagrams from hi-res data with multi-temperature fits of both galaxy nuclei and UGC 12914N and S (left panel) and the bridge/arm region (right panel). The fit parameters are shown in Table~\ref{tab:excipars}.}\label{fig:hiexcite}
\end{figure}

A two-temperature fit can also be derived for hi-res bridge/arm region, since the hi-res data includes the S(2) line. We find $T_1 = 102$~K for the coolest component, and $T_2 = 310$~K for the warmer one (Figure~\ref{fig:hiexcite}, right panel). However, with only three observed points there is significant degeneracy between the temperature and the column density, so the results should be taken as very approximate and poorly constrained. Adopting the above temperatures, the column density of $1.5\times10^7$~$M_\odot$~kpc$^{-2}$ in the cooler component would be 2.5 times the value derived from the low-res data, and is probably unrealistic. We consider them as strict upper limits only.

Table~\ref{tab:excipars} also tabulates the \Ht line fluxes for the 0--0~S(0)--S(2) lines for the nucleus and outer disk positions of UGC 12914, and the UGC~12915 nucleus as measured from the hi-res spectra. Figure~\ref{fig:hiexcite} shows the fits to the excitation diagrams for the positions on the galaxies in the left panel.

\section{Discussion} \label{sec:discussion}
\subsection{\Ht heating sources} \label{sec:heating}
\subsubsection{PDR heating} \label{sec:uv}

\begin{figure}
%\epsscale{.50}
\plotone{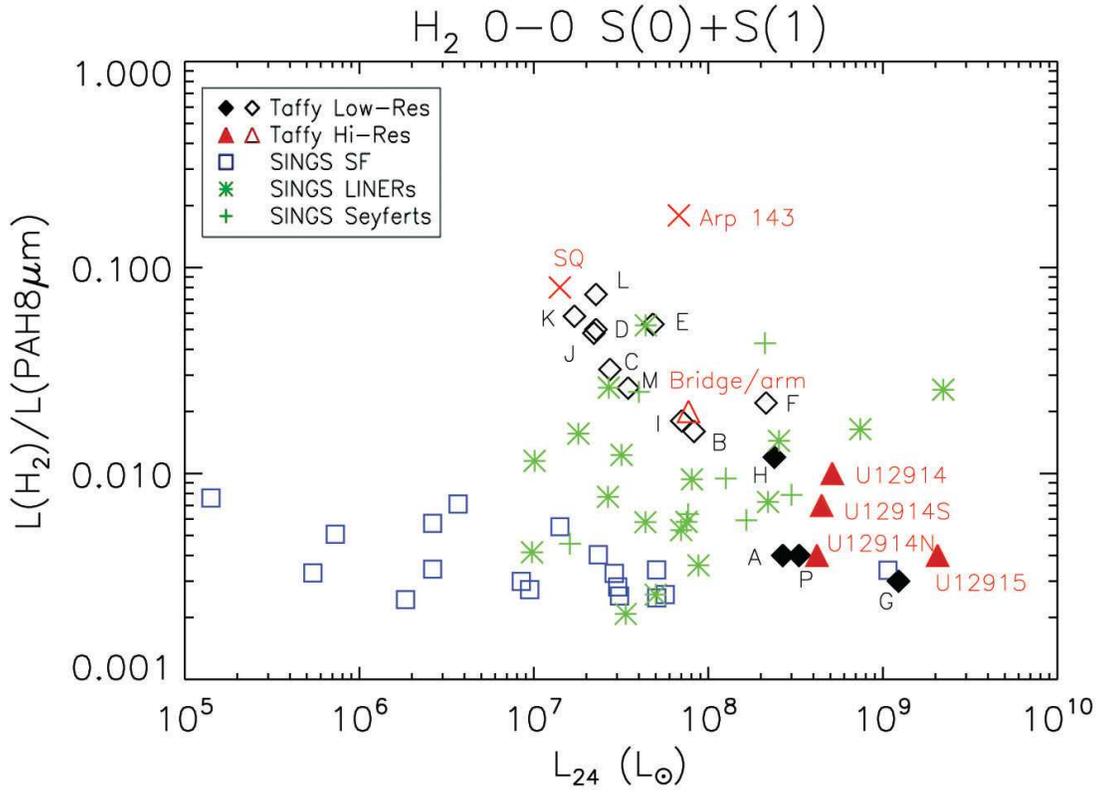}
\caption{Total luminosity in H$_2$ 0--0 S(0) and S(1) lines relative to PAH emission in the IRAC 8.0~$\micron$ band. Extraction regions which fall on the disks or their nuclei have filled symbols. The PAH emission was determined using rectangular apertures matched to the spectral extraction apertures in the IRAC 8.0~$\micron$ band, which was Gaussian-convolved to match the resolution of the MIPS 24~$\micron$ band and corrected for stellar contamination (see text). Luminosities at 24~$\micron$ were determined using MIPS 24~$\micron$ photometry, except for the high resolution Taffy apertures, for which an average of the spectrum over 22.5--25.0 \mic was used (see text). For comparison, we show galaxies from the SINGS sample \citep{rou07}, Arp 143 knot G \citep{bei09}, and the SQ shock sub-region \citep{clu10}.}\label{fig:h2pah}
\end{figure}

To quantify the strength of the \Ht emission, we compare the total flux in the \Ht 0--0 S(0) and S(1) lines with the PAH emission in the IRAC 8.0 $\micron$ band. The PAH8$\micron$ strength is estimated using the method of \citet{hel04}, in which the flux from the starlight-dominated 3.6~\mic band, scaled by a factor 0.232, is subtracted from the flux in the 8.0~$\micron$ band. Following \citet{ogl10}, we plot the luminosity ratio $L({\rm H_2})/L$(PAH8$\micron$) against the 24~$\micron$ luminosity $\nu L_{\nu}$, for both the low- and hi-res extraction regions in Figure~\ref{fig:h2pah}. The 24~$\micron$ luminosity is determined from the MIPS measurements for all of the data points, except for the hi-res apertures. For the hi-res apertures associated with the galaxies, the MIPS measurements are very sensitive to the exact centering of the extraction box, and so we self-consistently used the average flux over the range 22.5--25.0~$\micron$ from the spectrum to form the 24~$\micron$ continuum. This also ensures that any bias introduced in the extractions by assuming the lines come from a point source rather than an extended source do not somehow influence the results since the slit correction factor for point sources would disappear in the ratio. In addition to the Taffy regions, we show star forming galaxies, LINERS, and Seyferts from the SINGS sample \citep{rou07}. We also show the SQ shock sub-region \citep{clu10} and Arp 143 knot G, which had the highest $L({\rm H_2})/L$(PAH8$\micron$) ratio of any part of the Arp 143 system \citep{bei09}. $F$(PAH8$\micron$) has been determined in the same way for all data points.

In Figure~\ref{fig:h2pah} we have distinguished between spectra taken on the bright parts of the galaxy disks and spectra taken between the galaxies in the bridge. The spectra centered close to bright disk regions (or the nuclei themselves) show $L({\rm  H_2})/L$(PAH8$\micron$) ratios typical of star-forming galaxies and likely associated with photodissociation regions (PDRs) around young stars. The $L({\rm H_2})/L$(PAH8$\micron$) ratio in the Taffy bridge regions are much larger than in the bright disks, and are generally higher than those of the SINGS AGNs. We note that Region F, which includes the giant \ion{H}{2} region, shows a higher $L({\rm H_2})/L$(PAH8$\micron$) ratio than is typical of star forming galaxies. This is not surprising, since it lies in the bridge and may have both bridge and PDR-type heating present along the line of sight.
%The star forming galaxies indicate the normal $L({\rm H_2})/L$(PAH7.7$\micron$) ratio associated with UV-excited photodissociation regions (PDRs) around young stars. The $L({\rm H_2})/L$(PAH7.7$\micron$) ratio in the Taffy bridge regions are much larger than in these galaxies, and are generally higher than those of the AGNs, which suggests that the emission is not related to PDRs. Region F, which includes the giant \ion{H}{2} region, shows a higher $L({\rm H_2})/L$(PAH7.7$\micron$) ratio than is typical of star forming galaxies. This is not surprising, since it lies in the bridge and is thus observed through a column of warm \Ht gas.

There is generally good agreement between the high and low resolution data in the emission regions within the galaxies. The hi-res spectra of the UGC 12915 nucleus cover the same part of the system as region G, the UGC 12914 nucleus is comparable to region H, the northern clump of UGC 12914 is comparable to region A, and the UGC 12914 southern knot is not far from region P. These regions have $L({\rm H_2})/L$(PAH8$\micron$) ratios on the high end of the star forming galaxy distribution which, along with their positions within the galaxies, suggests that the excitation mechanism may be the same as that of the star forming (not AGN dominated) SINGS galaxies, namely PDR heating.
 
A similar conclusion is reached by examining Figure~\ref{fig:h2L24}, which shows the ratio $L({\rm H_2})/L_{24}$, as in \citet{ogl10}. In addition to the SQ shock sub-region and Arp 143 knot G, we also plot here 3C 326 N \citep{ogl07}. Emission at 24 \mic is mostly due to warm dust heated by young stars. The relative \Ht emission in the Taffy bridge regions is considerably stronger than that in the star forming galaxies, and is also stronger than most of the AGNs. The bridge regions are generally comparable to 3C 326 N and Arp 143 knot G, but somewhat weaker than the SQ shock sub-region. 

The possibility of PDR heating can also be investigated by comparing to the Meudon PDR models of \citet{lep06}. The models use the flux ratio of the \Ht S(0)--S(1) lines to the CO(1--0) line to measure the heating rate in the molecular gas \citep{nes10, gui12}. We use the parameters of \citet{hab11} to produce four different models, with $n_{\rm H}=100$ and 1000~cm$^{-3}$ and UV radiation scaling factor $G_{\rm UV}=1$ and 10. The radiation scaling factor is defined relative to the UV field of \citet{hab68}. The models predict $F({\rm H_2})/F({\rm CO})\approx20$--60 for PDR heating. For comparison with the models, we use the CO(1--0) data of \citet{gao03}, which give $F({\rm CO}) =1.2\times10^{-18}$~W~m$^{-2}$ over the entire bridge. Using the total \Ht bridge flux of $5.8\times10^{-16}$~W~m$^{-2}$, we find $F({\rm H_2})/F({\rm CO})=460$, about an order of magnitude above the predicted levels for PDR excitation.

The two galaxies are not exceptionally strong \Ht emitters compared to the SINGS star forming galaxies, with $L({\rm H_2})/L$(PAH8$\micron$) ratios only slightly above the most ${\rm H_{2}}$-bright star forming galaxies. In contrast to most of the bridge, the two galaxies both host star formation. UGC 12915 seems to be in an early starburst phase \citep{jar99}, so much of the \Ht emission is likely to have a PDR origin. Star formation in UGC 12914 does not appear to be as strongly enhanced \citep{jar99}, but the \Ht emission is consistent with star forming galaxies.

For comparison with the bridge, we compare the $F({\rm H_2})/F({\rm CO})$ ratios of  UGC 12915 nucleus (region G) and UGC 12914 South (region P). Using the CO(1--0) line profiles of \citet{gao03} and rescaling to match the areas of bridge region apertures G and P, we find CO fluxes $F({\rm CO})=9.4\times10^{-19}$ and $3.3\times10^{-19}$~W~m$^{-2}$. This gives $F({\rm H_2})/F({\rm CO})=78$ and 98 in UGC 12915 and UGC 12914 South, respectively. Both galaxies have stronger \Ht emission than expected for PDR heating, but are within a factor of two of the models. It seems plausible that PDR heating is a significant source of heating in the galaxies, but there are probably additional heating sources as well. More significantly, the line ratios in the galaxies are considerably smaller than that in the bridge, highlighting the likely difference in the dominant heating mechanism.

\subsubsection{Cosmic ray heating}
Given the relatively strong radio emission from the bridge region, we consider whether cosmic ray (CR) heating could be responsible for the excitation of the warm \Ht emission. This process involves low-energy cosmic rays (1--10~MeV), which ionize some of the gas. The resulting primary and secondary electrons heat the gas, which excites the \Ht through collisions \citep{dal99}. Radio continuum observations of the bridge provide some clues as to the energy density of cosmic rays. We caution however that the CRs that are expected to excite \Ht are at the very low energy tail of the CRs that give rise to the radio continuum emission.

The VLA 20~cm radio continuum flux in the mid-bridge is $\sim1.1$~mJy per 5$\arcsec$ beam \citep{con93}. We assume a power law spectral distribution of the form $\nu^{-\alpha}$ with $\alpha=1.2$ over the range 1.49--4.96~GHz, and a plasma depth of 12~kpc (roughly the scale of the bridge). The minimum equipartition magnetic field is found to be 6.4~$\mu$G \citep[e.g.,][]{gov04} for a lower spectral cut-off at 10~MHz. This is close to the value obtained by \citet{con93} under similar assumptions.  The corresponding magnetic energy density is $3.8\times10^{-13}$~J~m$^{-3}$. This energy density is uncertain to within at least a factor of two because of uncertainties in the value of the proton/electron energy ratio (assumed unity), as well as uncertainties in the volume filling factor of the magnetic field (also assumed to be unity) and the depth of the plasma column.

If the energy density in the cosmic rays $<U_{\rm CR}>$ $\sim$ $<U_B>_{\rm min}$, then we can estimate the power available for
heating $P_{\rm heat}$ = $<U_{\rm CR}>$ $\times$ $\eta/\tau$, where $\eta$ is the efficiency of the CR heating of ${\rm H_2}$, and $\tau$ is the characteristic deposition timescale. For the central bridge, assuming a depth of 12~kpc, this corresponds to a luminosity surface density $L_{\rm CR}$ = 4.5 $\times$ 10$^{32}$ ($\eta$/$\tau_7$)~W~kpc$^{-2}$, where $\tau_7$ is the deposition timescale in units of 10~Myr, the approximate expansion timescale of the bridge. We estimate $\eta$ by approximating the energy deposited by a typical MeV cosmic ray from its specific stopping power measured in MeV per unit column density. A value of 3.5~MeV g$^{-1}$ cm$^{-2}$ is quoted for typical MeV cosmic rays in the ISM \citep{yus07}, where the column density is that of the target material, in this case a mixture of molecular hydrogen and \ion{H}{1}. With \Ht column densities in the Taffy bridge of $\sim$$2\times10^{20}$ molecule cm$^{-2}$, the stopping power liberated by the passage of these CRs through this medium would be of order $3.5\times10^{-3}$~MeV per particle. Even allowing for the existence of more cold \Ht material, this implies $\eta \lesssim 0.01$. In order to power the \Ht for the lifetime of the bridge ($\tau_7=1$) with a measured \Ht surface brightness  of $\sim$$1.5 \times10^{32}$~W~kpc$^{-2}$, a cosmic ray energy density of at least 100 times that derived from equipartition arguments would be required. Alternatively, CRs could heat the \Ht if we are observing the system during a deposition burst ($\tau_7 \sim$ 0.01). Such a process would have to be global, perhaps taking the form of a sudden injection of CRs streaming from the galactic disks, but this seems unlikely.

Another approach to evaluating the influence of cosmic rays, which is independent of assumptions about equipartition, is to estimate the ionization rate $\zeta_{\rm CR}$ needed to balance the \Ht line cooling if CRs were the primary heating source. The adopted \Ht surface brightness of 1.5$\times10^{32}$~W~kpc$^{-2}$ gives a cooling rate $\sim5\times 10^{-32}$~W~molecule$^{-1}$, assuming an average mass density of $5.3 \times10^6$~$M_{\odot}$~kpc$^{-2}$ over the bridge. The heating rate by CR ionization has been estimated by various authors. \citet{yus07} estimate the CR heating to be $2 \times (4 \times10^{-18}\zeta_{\rm H_2})$~W~molecule$^{-1}$, where $\zeta_{\rm H_2}$ is the \Ht ionization rate. Balancing \Ht cooling with CR heating under these assumptions would require an ionization rate of  $\sim$10$^{-14}$~s$^{-1}$. Under the same assumptions, \citet{ogl10} and \citet{nes10} infer similar values for the ionization rate in the MOHEG radio galaxy 3C 326, where shocks were implicated.

This ionization rate is significantly higher than that measured in the molecular clouds in the Galactic Center \citep{oka05}. Thus, on purely comparative grounds, a CR ionization rate in the Taffy bridge would again require an unusually high CR energy density, perhaps 10 times that of the unusual Galactic Center regions and $\sim$100 times the galactic neighborhood. Based on these arguments we can conclude that cosmic rays are unlikely to be responsible for heating the \Ht in the Taffy bridge.

\subsubsection{Heating by magnetic reconnection}
This process is closely related to cosmic ray heating. In the previous subsection we estimated that the magnetic energy density in the bridge plasma was about $3.8\times10^{-13}$~J~m$^{-3}$. Again assuming a bridge depth of about 12~kpc, the energy column density is $\sim1.4\times10^{47}$~J~kpc$^{-2}$. Supposing that this magnetic energy is extracted on the bridge expansion timescale of about 10$^7$~yr \citep{con93}, we expect a surface luminosity of about, $F_{\rm rec} = 4.4\times10^{32}\tau_7$~W~kpc$^{-2}$.

This is comparable to the corresponding cosmic ray surface luminosity, as would be expected from the equipartition approximation of the previous subsection. Like the cosmic ray luminosity, this estimate must be corrected for an efficiency factor, representing the fraction of the reconnection energy used to excite ${\rm H_{2}}$. This should include both direct excitation from the radiation from reconnection regions, and indirect processes, like broader ambient heating from reconnection. In either case we expect a very low efficiency, since the plasma will be transparent to most wavebands of the broad radiation spectrum produced directly or indirectly in reconnection events. 

Note that the surface luminosity estimate above assumes the extraction of all of the magnetic energy on the adopted timescale, and so is an overestimate. On the other hand, it does not account for additional field generation in turbulent dynamos. This process may be locally important, but it is hard to see how it could have a global effect in the short time available. With the magnetic field estimate above, and mass density $10^{7}$~$M_{\odot}$~kpc$^{-2}$ estimated from \citet{gao03}, the Alfv\'{e}n velocity $v_{\rm A}\sim20$~km~s$^{-1}$, an order of magnitude less than the bridge expansion velocity of 450~km~s$^{-1}$ \citep{con93}, so even the propagation of magnetic effects must be very localized. To create significant power over the whole bridge would require some organized triggering of many localized reconnection events. One possible mechanism is large-scale turbulence, to be discussed in the next section. Turbulence could, in principal, create locally tangled magnetic fields which could then be stretched by the galaxies as they move apart, creating the conditions for possible reconnection. If the turbulence continued to be present over a long period of time, new events could continue to be created which might heat the ${\rm H_2}$. Thus, although unlikely, we cannot completely rule out magnetic reconnection as another source of \Ht heating.

Specifically, if the typical reconnection scale relative to the bridge size is much less than the ratio of the Alfv\'{e}n speed to the bridge expansion speed, then the reconnection timescale is much less than the age of the bridge. In that case, with a short timescale for reconnection, the reconnection luminosity is likely to decay on a similar timescale, yielding less luminosity than at earlier times.

\subsubsection{Turbulence and/or shock heating}
A more likely source for \Ht heating is turbulence, created in the wake of the collision between the two galaxies. We estimate turbulent heating rate $\Gamma_{\rm turb}$ using

\begin{displaymath}
\Gamma_{\rm turb} = \frac{3}{2}\times\frac{2m_{\rm H}(v/2.36)^2}{t_{\rm dis}},
\end{displaymath}

where $v$ is the velocity width over a length scale $l$, the factor 2.36 converts the line width to the rms velocity of the gas, and $t_{\rm dis} = l/(v/2.36)$ is the energy dissipation timescale \citep[e.g.,][]{mac99, tie05}. From the CO observations of \citet{gao03}, we take $v \approx 200$~km~s$^{-1}$ over a beam of 14$\arcsec$, or $l\approx4$ kpc. This yields $t_{\rm dis}\approx5\times10^{7}$~yr. The heating rate is then $\Gamma_{\rm turb}\approx 2.5\times10^{-32}$~W~molecule$^{-1}$, about 50\% of the observed cooling rate of $5\times10^{-32}$~W~molecule$^{-1}$. Turbulence on this scale could contribute significantly to the gas heating, provided that it is relatively efficient.

A related possibility is heating by shocks. As noted earlier, the H$_2$ line cooling time is very short, and {\it in-situ} heating by shocks or turbulence is consistent with the need to maintain the observed gas at $\sim$160--170~K for a reasonable fraction of the bridge lifetime.

We may obtain a crude estimate of the energy available in shocks by examining the bulk mechanical energy associated with the post-shock gas. The CO velocity dispersion, which should roughly track the velocity dispersion of the cold {\rm H$_2$}, is (200/2.36)~km~s$^{-1}$ \citep{gao03}. \citet{zhu07} found the \Ht mass to be $M_{\rm H_2}\sim1.3\times10^{9}$~$M_{\odot}$, so over the 20~Myr since the collision the average heating rate is $1.5\times10^{34}$~W. This is comparable to the bridge \Ht luminosity of $2.5\times10^{34}$~W.

This estimate can be checked against an upper limit to the available power estimated using the bulk kinetic energy of the \ion{H}{1} and \Ht gas in the bridge. A re-analysis of the radio data of \citet{con93} by \citet{gao03} determined that the \ion{H}{1} mass in the bridge is $M_{\rm HI}\sim6\times10^{9}$~$M_{\odot}$. We take this mass as the pre-shock \ion{H}{1} mass, and a collision velocity of 450~km~s$^{-1}$ \citep{con93} in the shock frame. The time since the collision is $\sim20$~Myr, so the total available power from the \ion{H}{1} is $2\times10^{36}$~W, about 2 orders of magnitude higher than the \Ht luminosity.

Of course, not all of the available power from any heating mechanism will be dissipated via \Ht line emission. We note that H$_2$ will not be the only line-coolant in the shock, although it could be a significant one. For example, in our modeling of the similar shock-heating of the Stephan's Quintet system, we estimate that 75\% of the line cooling in the molecular shocks is from H$_2$, with the majority of the other cooling through H$_2$O, CO and [\ion{O}{1}]$\micron$ line emission. The details of the cooling channel would depend somewhat on the mix of C- and J-shocks (see \citealt{flower10}).

Other, higher velocity shocks which might be present  could also dissipate significant energy through mechanisms such as Ly-$\alpha$ emission and other UV coolants, as well as diffuse X-ray gas heated in the collision. There is not currently any X-ray data available for the Taffy that could be used to determine the significance of X-ray cooling in the system. In Stephan's Quintet, the total X-ray luminosity is about 3 times lower than the \Ht line luminosity \citep{clu10}, but it is not clear that the same fraction should apply to the Taffy. The presence of shocks is required by the recent model of \citet{lis10}. In the model, strong shocks in the Taffy bridge accelerate charged particles to ultrarelativistic energies, which then produce the synchrotron emission observed in the radio continuum. 

We conclude that the largest heating source is probably shocks, as in the model presented by \citet{gui09} which was developed to explain the powerful \Ht emission detected in the giant intergalactic shock wave in Stephan's Quintet. The essential ingredients of this model are a high-speed shock wave that is driven into a multi-phase medium, leading to the collapse and shock-heating of denser material. Molecules form rapidly on dust grains which survive in the denser clumps of material, and strong \Ht emission results. Models fitted to the \Ht excitation diagrams in Stephan's Quintet show that the emission is consistent with several low and medium velocity C-shocks, although more destructive J-shocks cannot be ruled out. Our observations do not provide enough points on the excitation diagram to justify extensive shock-modeling, but the similarities with the spectra seen in the Taffy bridge (powerful \Ht with only low-excitation metal lines) and the approximately similar low-$J$ temperature fits suggest a common origin for the \Ht excitation. 

\begin{figure}
%\epsscale{.70}
\plotone{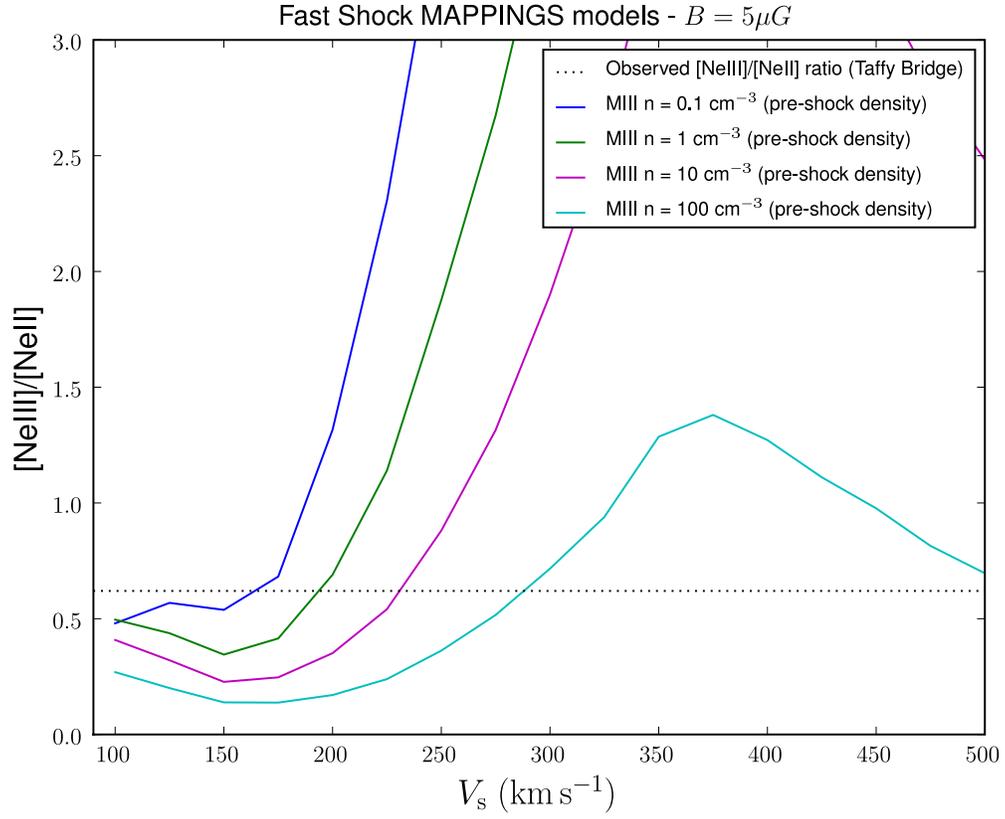}
\caption{MAPPINGS shock velocity models \citep{all08} for magnetic field $B=5\mu$G. The dotted line indicates [\ion{Ne}{3}]/[\ion{Ne}{2}] = 0.62, as measured in the bridge using the IRS SH module. The curves correspond to different values for the pre-shock density.}
\label{fig:shockz}
\end{figure}

By analogy with Stephan's Quintet, a distribution of shock velocities, related to the density distribution in the shocked gas, is likely to be present \citep{gui09}. In the gas that is less dense than the molecular gas, shock velocities $>$ 100 km s$^{-1}$ are likely, which will ionize the medium. These shock velocities can be estimated using the observed ratio [\ion{Ne}{3}]15.56 $\micron$/[\ion{Ne}{2}]12.81~$\micron$, which we obtain from the hi-res bridge/arm region where both of these lines were observed with the SH module.\footnote{The [\ion{Ne}{2}]12.81~$\micron$ was observed in the low-res data only in regions C and J. Since it falls on a separate module from the [\ion{Ne}{3}]15.56 $\micron$ line, the ratio will depend on the scaling factor chosen between the two modules. We therefore opt instead to use the hi-res bridge/arm region, where both lines fall on the SH module and the ratio is independent of scaling.} These lines can also originate with star formation, and the observed line ratio [\ion{Ne}{3}]/[\ion{Ne}{2}] $\approx0.62$ in the bridge/arm region is within the range typical of star forming regions \citep{dale06}, but H$\alpha$ maps show no evidence of significant star formation in the Taffy bridge \citep{bus90}, though star formation is likely to be present in the UGC 12914 spiral arm. We use the MAPPINGS shock models of \citet{all08} for a magnetic field of 5 $\mu$G and a pre-shock gas density between 0.1--100~cm$^{-3}$, as shown in Figure~\ref{fig:shockz}. The shock velocity for the Taffy bridge falls in the range 100--300~km~s$^{-1}$, comparable to the SQ shock \citep{clu10}.

\begin{figure}
%\epsscale{.50}
\plotone{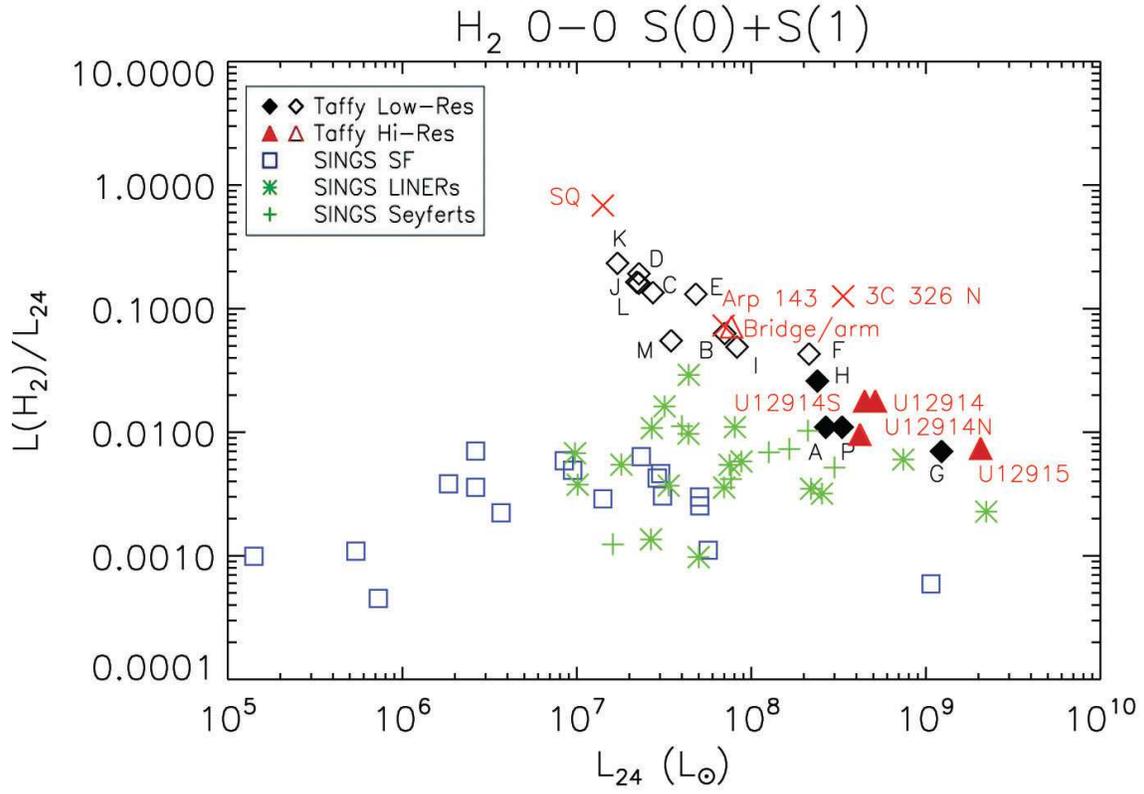}
\caption{Total luminosity in H$_2$ 0--0 S(0) and S(1) lines relative to 24~\mic emission. Other objects are shown as in Figure~\ref{fig:h2pah}. In addition, we include 3C 326N \citep{ogl07}.}\label{fig:h2L24}
\end{figure}

Figures~\ref{fig:h2pah} and \ref{fig:h2L24} may provide some insight into the location of the strongest shocks. Figure~\ref{fig:h2pah} shows a spatial trend in $L({\rm H_2})/L$(PAH), with values close to those of the galaxies at the bridge/galaxy interface growing to become comparable to the SQ shock near the center of the bridge. A similar trend is found in $L({\rm H_2})/L_{24}$, shown in Figure~\ref{fig:h2L24}. This suggests that the shocks in the Taffy bridge are strongest relative to other excitation mechanisms near its center.

\subsection{Additional comparison with other systems}

It is worth making some additional comparison between the properties of the Taffy bridge and those in the large-scale shocked filaments in the Stephan's Quintet system \citep{app06, clu10}. The mean surface brightness of the low resolution S(0) and S(1) lines is $\sim$1.5$\times10^{32}$~W~kpc$^{-2}$, more than a factor of 2 larger than the equivalent emission averaged over the inner region of the main shock in Stephan's Quintet, which we estimate to be 7$\times$10$^{31}$~W~kpc$^{-2}$ for the S(0) and S(1) lines. Thus the surface brightness in the Taffy is greater, reflecting a larger surface mass density of excited gas. We can probably rule out higher excitation of the warm gas as the explanation because, as we show in Section~\ref{sec:CO}, the ratio of warm to total \Ht gas in the Taffy is probably lower (within the considerable uncertainties) than that in SQ. Thus the higher surface luminosity is almost certainly a result of the different quantities of gas being swept up in the shock in the Taffy compared with SQ, because of the different nature of the collisions in the two systems.

Alternatively, the higher surface density of warm gas may be purely geometrical -- the Taffy bridge could be deeper in the plane of the sky than the SQ shock, a result of very different collision geometries in the two cases. The total gas mass in the Taffy bridge would depend on many complicated factors resulting from the geometry of the original collision, including disk impact parameters and inclination and the rotational kinematics of the disks. All of these factors would govern how much mass is splashed into the bridge and its projected surface density (see \citealt{str97}).

Local Luminous Infrared Galaxies (LIRGs) and Ultra-luminous Infrared Galaxies (ULIRGs) comprise a high fraction of collisional or merging systems. The Taffy system has a far-IR luminosity as measured by \citet{san03} $L_{\rm IR}=10^{10.9}$~$L_{\odot}$, just short of the definition of a LIRG. However, given the head-on collision between the Taffy galaxies, it is likely that they will eventually merge through dynamical friction. Thus the Taffy may move from its current pre-LIRG status to that of a LIRG or ULIRG if the two galaxies are drawn back together. Extended mid-IR \Ht emission is seen in $\sim$70$\%$ of a sample of nearby ULIRGs \citep{hig06}, and shocks driven into their outer disks have been hypothesized but not proven \citep{zak10}. The existence of extended shocked \Ht in the Taffy system 20~Myr after the initial collision suggests that \Ht emission can persist for a time comparable with the dynamical crossing time. Thus given the even more chaotic nature of the gas-rich major mergers believed to explain many local ULIRGs, and their expected rapid coalescence rates, it seems plausible that the \Ht emission seen in ULIRGs may result from shocked processes broadly similar to that seen in the Taffy.

\subsection{Warm \Ht gas fraction}
\label{sec:CO}
One of the most remarkable features of the Taffy system is the fact that the peak \ion{H}{1} concentration is in the bridge between the two galaxies \citep{con93}. In contrast, the CO~(1--0) map of \citet{gao03} shows that the molecular gas has its highest concentrations in the galaxies. However, the molecular gas is present throughout the bridge, so that the total \Ht mass in the bridge appears to be comparable to that of the galaxies.

The estimated total \Ht mass depends on the conversion factor $X = N_{\rm H_2}/I_{\rm CO}$, which is unknown. To estimate the warm gas fraction in the bridge, defined as the ratio of the warm \Ht mass derived from the low-res IRS observations to the \Ht mass derived from the CO observations, we first use the Galactic value of $X=2\times10^{20}$~cm$^{-2}$/(~K~km~s$^{-1}$). The warm \Ht mass is estimated to be $8\times10^8$~$M_{\odot}$ based on the S(0) and S(1) lines (see Section \ref{sec:exc}). This gives a a warm gas fraction in the bridge of 0.09. We note that in the case of Stephan's Quintet, recent measurements with the IRAM 30~m telescope \citep{gui12} suggest comparable values along the main intergalactic shock $M_{\rm H_2, warm}/M_{\rm H_2} = 0.10$--0.33 using the same Galactic $X$ factor.

%We first use the Galactic value of $X=2\times10^{20}$~cm$^{-2}$/~(K~km~s$^{-1}$) to estimate the warm gas fraction in the bridge, defined as the ratio of the warm \Ht mass derived from the low-res IRS observations to the \Ht mass derived from the CO observations. Assuming this Galactic $X$ factor, we find a warm gas fraction in the bridge of 0.09. We note that in the case of Stephan's Quintet, recent measurements with the IRAM 30~m telescope \citep{gui12} suggest comparable values along the main intergalactic shock $M_{\rm H_2, warm}/M_{\rm H_2}$ = 0.10--0.33 using the same Galactic $X$ factor.

However, the value of the $X$-factor in the Taffy bridge could be lower than the Galactic value by a factor of a few. Based on $^{13}$CO measurements, \citet{bra03} suggested that the $X$ factor could be at least a factor of 4 lower than the Galactic value in the bridge region, which would increase the warm gas fraction by a factor of four.

\citet{zhu07} fit large velocity gradient models to CO line ratios and determined that $X = 2.6\times10^{19}$~cm$^{-2}$/(K km s$^{-1}$) in the bridge, which implies $M_{\rm H_2} = 1.3\times10^{9}$~$M_{\odot}$ and $M_{\rm H_2, warm}/M_{\rm H_2}=0.7$. Therefore, the warm \Ht gas fraction in the bridge is poorly constrained, likely being in the range $0.09-0.7$. This is on the upper end of the distribution for the SINGS galaxies \citep{rou07}. Together with the absence of obvious source of UV heating, the bracketed range of warm \Ht gas fraction suggests an active heat source from shocks.

%NOTE: only ours is for s(0) and s(1) only

\section{Summary and Future Work} \label{sec:summary}
We have reported on the detection of powerful mid-IR emission lines of \Ht in the Taffy galaxies. Our main results are:

\begin{enumerate}
 
\item The mid-IR spectrum of the Taffy bridge bears a striking similarity to the spectrum of the group-wide shock in Stephan's Quintet. In particular, the  $L({\rm H_2})/L$(PAH8$\micron$) ratio is unusually high in the bridge, exceeding by an order of magnitude that found in SINGS star forming galaxies \citep{rou07}, while the galaxies UGC 12914/5 are comparable to star forming galaxies. 
\item The warm \Ht is distributed throughout the bridge between the two galaxies. This is similar to the \Ht detected previously using sub-mm observations of CO. It is also comparable to the distribution of 20~cm radio continuum and dust emission.
\item Single-temperature fits to the two \Ht lines measured throughout the bridge indicate a nearly uniform temperature of 150--175~K. We also estimate a mass density $\sim$$5\times10^6$~$M_{\odot}$~kpc$^{-2}$. We measure a warm \Ht mass of $4.2\times10^8$~$M_{\odot}$ in the observed parts of the bridge, and predict a total mass in the complete bridge of 9(+2/$-5)\times10^8$~$M_{\odot}$ if the rest of the bridge has the same properties as that observed.
\item The \Ht surface luminosity in the Taffy bridge is more than a factor of two larger than the main shock region in SQ, probably 
because of different collision geometries and initial conditions. Since the \Ht line cooling time is many orders of magnitude shorter than the dynamical timescale for the bridge, {\it in situ} heating is required to explain the observed warm temperature of the bridge gas. The \Ht gas is probably heated by turbulence and shocks  produced during the collision, and also by continuing cloud collisions in the bridge.
%\item The \Ht gas is probably heated by shocks produced during the collision, similar to the shock in Stephan's Quintet, and also by continuing cloud collisions in the bridge. The \Ht surface luminosity in the Taffy bridge is more than a factor of two larger than the main shock region in SQ, probably because of different collision geometries and initial conditions.
\end{enumerate}

We have determined that the \Ht molecule provides an important cooling channel for the warm gas in the Taffy bridge. Other molecules may also be important. Future observations (granted in Open Time 1) with the \emph{Herschel Space Observatory} will allow us to further explore the thermodynamic state of the shocked ISM bridge.

\acknowledgments

BWP thanks the IPAC Visiting Graduate Student Fellowship (from September 2009 to March 2010) for supporting this work. We also thank Jim Condon for providing a digital copy of the 20~cm radio data and the anonymous referee for many insightful comments which greatly improved this manuscript. This work is based on observations made with the \emph{Spitzer Space Telescope}, which is operated by the Jet Propulsion Laboratory (JPL), California Institute of Technology (Caltech) under NASA contract. This research has made use of the NASA/IPAC Extragalactic Database (NED) which is operated by the Jet Propulsion Laboratory, California Institute of Technology, under contract with the National Aeronautics and Space Administration.

{\it Facilities:} \facility{Spitzer}.

\clearpage

\begin{deluxetable}{lrrccccc}
\tablecolumns{8}
\tablewidth{0pc}
\tablecaption{H$_2$ Line Fluxes ($10^{-17}$ W m$^{-2}$) \label{tab:h2}}
\tablehead{
\colhead{Region} & 
\colhead{RA} &
\colhead{Dec} &
\colhead{H$_2$ 0--0 S(0)} & 
\colhead{H$_2$ 0--0 S(1)} & 
\colhead{H$_2$ 0--0 S(2)} &
\colhead{H$_2$ 0--0 S(3)} &
\colhead{Scaling\tablenotemark{a}}\\
&
\colhead{(J2000.0)} & 
\colhead{(J2000.0)} &
\colhead{$\lambda28.22\micron$} & 
\colhead{$\lambda17.03\micron$} & 
\colhead{$\lambda12.28\micron$} &
\colhead{$\lambda9.66\micron$} &
\colhead{}\\
}
\startdata
\cutinhead{High Resolution\tablenotemark{b}}
UGC 12915 nucleus & 0 01 41.89 & +23 29 44.0 & 3.17 $\pm$ 0.25 & 10.41 $\pm$ 0.19 & 3.02 $\pm$ 0.09 & \nodata & SH; 1.42\\
UGC 12914 nucleus & 0 01 38.09 & +23 29 03.3 & 1.57 $\pm$ 0.21 & 6.57 $\pm$ 0.14 & 2.30 $\pm$ 0.10 & \nodata & SH; 1.35\\
UGC 12914 S & 0 01 39.02 & +23 28 43.0 & 2.19 $\pm$ 0.27 & 4.89 $\pm$ 0.10 & 1.51 $\pm$ 0.08 & \nodata & SH; 1.42\\
UGC 12914 N & 0 01 37.37 & +23 29 18.5 & 1.06 $\pm$ 0.13 & 2.53 $\pm$ 0.12 & 0.72 $\pm$ 0.10 & \nodata & SH; 1.93\\
bridge/arm & 0 01 38.89 & +23 29 30.0 & 1.26 $\pm$ 0.13 & 3.61 $\pm$ 0.09 & 1.25 $\pm$ 0.10 & \nodata & SH; 2.81\\
\cutinhead{Low Resolution\tablenotemark{c}}
A & 0 01 37.62 & +23 29 18.7 & 0.50 $\pm$ 0.16 & 2.10 $\pm$ 0.24 & \nodata & \nodata & \nodata\\
B & 0 01 38.29 & +23 29 22.9 & 0.71 $\pm$ 0.08 & 3.21 $\pm$ 0.27 & \nodata & \nodata & \nodata\\
C & 0 01 38.97 & +23 29 27.0 & 0.63 $\pm$ 0.04 & 2.65 $\pm$ 0.16 & \nodata & \nodata & \nodata\\
C (PAHFIT)\tablenotemark{d} & 0 01 38.97 & +23 29 27.0 & 0.61 $\pm$ 0.09 & 2.74 $\pm$ 0.20 & 0.40$\pm$0.03 & 0.61$\pm$0.04 & SL1, SL2; 3\\
D & 0 01 39.64 & +23 29 31.1 & 0.68 $\pm$ 0.08 & 3.18 $\pm$ 0.17 & \nodata & \nodata & \nodata\\
E & 0 01 40.32 & +23 29 35.2 & 0.95 $\pm$ 0.12 & 4.67 $\pm$ 0.16 & \nodata & \nodata & \nodata\\
F & 0 01 40.99 & +23 29 39.3 & 1.33 $\pm$ 0.03 & 6.88 $\pm$ 0.18 & \nodata & \nodata & \nodata\\
G & 0 01 42.00 & +23 29 45.5 & 1.03 $\pm$ 0.28 & 6.31 $\pm$ 0.22 & \nodata & \nodata & \nodata\\
H & 0 01 38.06 & +23 29 04.8 & 0.61 $\pm$ 0.16 & 4.94 $\pm$ 0.25 & \nodata & \nodata & \nodata\\
I & 0 01 38.74 & +23 29 08.9 & 0.46 $\pm$ 0.04 & 3.18 $\pm$ 0.12 & \nodata & \nodata & \nodata\\
J & 0 01 39.41 & +23 29 13.0 & 0.51 $\pm$ 0.09 & 2.72 $\pm$ 0.13 & \nodata & \nodata & \nodata\\
J (PAHFIT)\tablenotemark{d} & 0 01 39.41 & +23 29 13.0 & 0.72 $\pm$ 0.06 & 2.90 $\pm$ 0.16 & 0.37$\pm$0.02 & 0.56$\pm$0.03 & SL1, SL2; 3\\
K & 0 01 40.09 & +23 29 17.1 & 0.72 $\pm$ 0.15 & 2.83 $\pm$ 0.21 & \nodata & \nodata & \nodata\\
L & 0 01 40.77 & +23 29 21.3 & 0.48 $\pm$ 0.08 & 2.78 $\pm$ 0.21 & \nodata & \nodata & \nodata\\
M & 0 01 41.44 & +23 29 25.4 & 0.30 $\pm$ 0.06 & 1.41 $\pm$ 0.12 & \nodata & \nodata & \nodata\\
N & 0 01 42.12 & +23 29 29.5 & $<0.41$\tablenotemark{e}   & 1.02 $\pm$ 0.22 & \nodata & \nodata & \nodata\\
O & 0 01 42.79 & +23 29 33.6 & $<0.25$\tablenotemark{e}   & 0.65 $\pm$ 0.12 & \nodata & \nodata & \nodata\\
P & 0 01 39.15 & +23 28 43.6 & 0.65 $\pm$ 0.18 & 2.62 $\pm$ 0.21 & \nodata & \nodata & \nodata\\
Q & 0 01 40.16 & +23 28 49.8 & 0.50 $\pm$ 0.13 & 0.86 $\pm$ 0.16 & \nodata & \nodata & \nodata\\
R & 0 01 40.84 & +23 28 53.9 & 0.36 $\pm$ 0.10 & 0.69 $\pm$ 0.25 & \nodata & \nodata & \nodata\\
S & 0 01 41.51 & +23 28 58.0 & $<0.16$\tablenotemark{e}   & 0.93 $\pm$ 0.12 & \nodata & \nodata & \nodata\\
T & 0 01 39.60 & +23 28 29.7 & $<0.25$\tablenotemark{e}   & $<0.46$\tablenotemark{e}   & \nodata & \nodata & \nodata\\
U & 0 01 40.61 & +23 28 35.9 & $<0.27$\tablenotemark{e}   & $<0.46$\tablenotemark{e}   & \nodata & \nodata & \nodata\\
\enddata
\tablenotetext{a}{Due the different angular size of the IRS modules, some of the spectra were rescaled so that the continua matched. The module and scaling factor are indicated here.}
\tablenotetext{b}{H$_2$ S(0) measured with \emph{Spitzer} IRS LH module; S(1) and S(2) lines measured with IRS SH module. All fluxes were measured using SMART.}
\tablenotetext{c}{H$_2$ S(0) line measured in LL1, S(1) in LL2, S(2) and S(3) where available in SL1. Low resolution bridge region apertures were squares $10\farcs15\times10\farcs15$. Fluxes were measured using SMART except where otherwise noted.}
\tablenotetext{d}{Line fluxes measured by fitting the full spectrum using PAHFIT, which requires a smooth continuum. In region J, SL1 and SL2 were both scaled up by a factor of 3 due to their smaller area. Then SL1, SL2, and LL2 were scaled up by a factor 1.86 to match the LL1 continuum. Following the fitting with PAHFIT, the SL1, SL2, and LL2 were scaled down by 1.86. No rescaling was required in region C.}
\tablenotetext{e}{$3\sigma$ upper limit estimated from the RMS and expected line width.}

\end{deluxetable}

\begin{deluxetable}{lcccccc}
\tablecolumns{7}
\tablewidth{0pc}
\tablecaption{Fine structure line fluxes for high resolution data ($10^{-17}$ W m$^{-2}$)
\label{tab:fine}}
\tablehead{
\colhead{Region} & 
\colhead{[\ion{Ne}{2}]} & 
\colhead{[\ion{Ne}{3}]} & 
\colhead{[\ion{S}{3}]} &  
\colhead{[\ion{Fe}{2}]+[\ion{O}{4}]} & 
\colhead{[\ion{S}{3}]} & 
\colhead{[\ion{Si}{2}]} \\
&
\colhead{$\lambda12.81\micron$} & 
\colhead{$\lambda15.56\micron$} & 
\colhead{$\lambda18.71\micron$} &   
\colhead{$\lambda25.99\micron$+$\lambda25.89\micron$} & 
\colhead{$\lambda33.48\micron$} & 
\colhead{$\lambda34.82\micron$}\\
}
\startdata
UGC 12915 nuc & 27.94$\pm$1.08 & 3.79$\pm$0.18 & 10.25$\pm$0.45 & 1.91$\pm$0.53 & 18.72$\pm$1.61 & 28.06$\pm$0.68 \\
UGC 12914 nuc & 4.26$\pm$0.28 & 1.52$\pm$0.11 & 1.79$\pm$0.15 & 0.65$\pm$0.15 & 2.89$\pm$0.59 & 8.34$\pm$0.37 \\
UGC 12914 S & 9.14$\pm$0.39 & 1.42$\pm$0.07 & 5.96$\pm$0.17 & 0.33$\pm$0.10 & 9.95$\pm$0.73 & 10.78$\pm$0.43 \\
UGC 12914 N & 6.35$\pm$0.32 & 1.53$\pm$0.07 & 4.20$\pm$0.20 & 0.38$\pm$0.14 & 7.93$\pm$0.47 & 9.41$\pm$0.54 \\
bridge/arm & 0.58$\pm$0.06 & 0.36$\pm$0.14 & 0.68$\pm$0.07 & $<0.23$\tablenotemark{a} & 0.58$\pm$0.23 & 1.63$\pm$0.28 \\
\enddata
\tablecomments{[\ion{Ne}{2}], [\ion{Ne}{3}], and [\ion{S}{3}]$18.71\micron$ measured with IRS SH module; [\ion{Fe}{2}]+[\ion{O}{4}], [\ion{S}{3}]$33.48\micron$, and [\ion{Si}{2}] measured with IRS LH module.The SH line fluxes have been scaled up using the same factors as in Table~\ref{tab:h2}.}
\tablenotetext{a}{$3\sigma$ upper limit estimated from the RMS and expected line width.}
\end{deluxetable}

\begin{deluxetable}{lllll}
\tablecolumns{5}
\tablecaption{H$_2$ Properties of Regions A--R
\label{tab:excipars}}
\tablehead{
\colhead{Region} &
\colhead{Temperature\tablenotemark{a}} &
\colhead{Equilibrium o/p\tablenotemark{b}} &
\colhead{$N_{\rm H_2}$} &
\colhead{${\Sigma}_{\rm H_2}$} \\
\colhead{} &
\colhead{(K)} &
\colhead{} &
\colhead{(10$^{20}$molecules~cm$^{-2}$)} &
\colhead{($10^6$~$M_{\sun}$ kpc$^{-2}$)} \\
}
\startdata
A          & 160 ($\pm$10) & 2.6 ($\pm$0.1)  & 2.7 (+1.0/-0.9)     & 4.4 (+1.4/-1.2) \\ 
B          & 163 ($\pm$10) & 2.6 ($\pm$0.1)  & 3.7 (+1.1/-0.9)     & 5.9 (+1.5/-1.2) \\ 
C          & 160 ($\pm$5)  & 2.6 ($\pm$0.1) & 3.4 ($\pm$0.4)      & 5.5 (+0.8/-0.7) \\
C-multi $T_1$ & 160 ($\pm$5)  & 2.6 ($\pm$0.1) & 3.2 ($\pm$0.4)      & 5.0 (+0.8/-0.7) \\
C-multi $T_2$ & 488 (+10/-70)  & 3.0         & 0.010 (+0.01/-0.001)     & 0.020 (+0.020/-0.001) \\
D          & 164 (+8/-7)   & 2.6 ($\pm$0.1) & 3.5 (+0.8/-0.7)     & 5.6 (+1.4/-1.2) \\
E          & 167 ($\pm$2)  & 2.7 ($\pm$0.1) & 4.6 ($\pm$0.2)      & 7.5 ($\pm$0.3) \\
F          & 169 ($\pm$2)  & 2.5 ($\pm$0.1) & 6.4 (+0.4/-0.3)     & 10.2 (+0.6/-0.4) \\
G          & 178 ($\pm$13) & 2.7 ($\pm$0.1) & 4.5 (+2.2/-1.6)    & 7.2 (+2.8/-2.5) \\
H          & 195 ($\pm$12) & 2.9 ($\pm$0.1) & 2.2  (+1.0/-0.8)      & 3.6 (+1.7/-0.7) \\
I          & 185 ($\pm$5)  & 2.8 ($\pm$0.1) & 1.9 (+0.5/-0.4)    & 3.0 (+0.5/-0.4) \\
J          & 165 ($\pm$3)  & 2.6 ($\pm$0.1) & 2.6 (+0.4/-0.1)     & 4.2 (+2.0/-0.2) \\
J-multi $T_1$ & 155 ($\pm$3)  & 2.5 ($\pm$0.1) & 4.1 (+0.4/-0.1)     & 6.6 (+2.0/-0.2) \\
J-multi $T_2$ & 433 ($\pm$3)  & 3.0 & 0.020 ($\pm$0.001) & 0.020 (+0.001/-0.001) \\
K          & 157 ($\pm$10) & 2.6 ($\pm$0.1)  & 4.1 ($\pm$1.8)      & 6.6 (+2.7/-0.1) \\
L          & 175 ($\pm$11) & 2.7 ($\pm$0.2)  & 2.1 ($\pm$0.7)      & 3.4 (+1.2/-1.2) \\
M          & 165 ($\pm$11) & 2.6 ($\pm$0.1)  & 1.5 ($\pm$0.6)     & 2.4 (+1.1/-0.8) \\
N & \nodata & \nodata & \nodata & \nodata \\
O & \nodata & \nodata & \nodata & \nodata \\
P          & 158 ($\pm$13) & 2.6 ($\pm$0.2)  & 3.6 ($\pm$2.0)     & 5.8 (+3.2/-2.5) \\
Q          & 130 (+14/-11) & 2.2 ($\pm$0.2)  & 4.6 (+3.0/-2.0)         & 7.3 (+5.0/-3.0) \\
R          & 133 (+10/-12) & 2.3 (+0.1/-0.3) & 3.1 (+1.0/-0.2)       & 5.0 (+1.8/-1.0) \\
UGC 12915 nuc $T_1$ & 143 (+5/-6) & 2.4 ($\pm$0.1) & 22 (+4/-3 ) & 36 (+6/-5) \\
UGC 12915 nuc $T_2$ & 696 (+30/-23) & 3.0  & 0.07 ($\pm$0.01) & 0.11 ($\pm$0.02) \\
UGC 12914 nuc $T_1$ & 151 (+8/-6) & 2.5 ($\pm$0.1) & 9.6 (+3.0/-2.0) & 15 (+4/-3) \\
UGC 12914 nuc $T_2$ & 793 (+10/-6) & 3.0  & 0.040 ($\pm$0.005) & 0.060 ($\pm$0.008) \\
UGC 12914 N $T_1$ & 135 (+8/-7) & 2.3 ($\pm$0.1) & 8.6 (+3/-2) & 14 (+4/-3) \\
UGC 12914 N $T_2$ & 1500 ($\pm$10) & 3.0  & 0.010 ($\pm$0.005) & 0.020 ($\pm$0.008) \\
UGC 12914 S $T_1$ & 132 (+4/-3) & 2.8 ($\pm$0.1) & 19 (+4/-3) & 30 (+8/-7) \\
UGC 12914 S $T_2$ & 989 ($\pm$40) & 3.0  & 0.020 ($\pm$0.005 ) & 0.040 ($\pm$0.008) \\
Hi res bridge/arm single-temp fit & 157 (+23/-14) & 2.6 ($\pm$0.2) & 2.3 (+1.7/-1.3) & 3.6 (+3.0/-1.9) \\
Hi res bridge/arm 2-temp fit $T_1$ & 103 (+27/-3)  & 1.7 (+0.5/-0.2)& 10 (+3/-4) & 15 (+5/-6) \\
Hi res bridge/arm 2-temp fit $T_2$ & 310 ($\pm$30) & 3.0 & 0.12 (+0.5/-0.7) & 0.2 (+0.05/-0.12) \\
\enddata
\tablenotetext{a}{Uncertainties in all derived properties are formal uncertainties in fitting and do not include many possible systematic effects}
\tablenotetext{b}{Equilibrium ortho-to-para ratio for \Ht molecules is assumed. Note that this can be significantly less than 3 for $T< 300$~K, though at high temperatures o/p = 3 with no formal uncertainty apart from that in the temperature determination. For Deviations from thermal equilibrium might be possible in shocks and this would further add to the uncertainties in the derived properties.}
\end{deluxetable}

%% The following command ends your manuscript. LaTeX will ignore any text
%% that appears after it.


\begin{thebibliography}{}
\bibitem[Allen et al.(2008)]{all08} Allen, M. G., Groves, B. A., Dopita, M. A., Sutherland, R. S., \& Kewley, L. J. 2008, \apj, 178, 20 %MAPPINGS
\bibitem[Appleton et al.(2006)]{app06} Appleton, P. N., et al. 2006, \apj, 639, L51
\bibitem[Beir\~{a}o et al.(2009)]{bei09} Beir\~{a}o, P., Appleton, P. N., Brandl, B. R., Seibert, M., Jarrett, T., \& Houck, J. R. 2009, \apj, 693, 1650
\bibitem[Braine et al.(2003)]{bra03} Braine, J., Davoust, E., Zhu, M., Lisenfeld, U., Motch, C., \& Seaquist, E. R. 2003, \aap, 408, L13
\bibitem[Bushouse \& Werner(1990)]{bus90} Bushouse, H. A., \& Werner, M. W. 1990, \apj, 359, 72
\bibitem[Cluver et al.(2010)]{clu10} Cluver, M. E., et al. 2010, \apj, 710, 248
\bibitem[Condon et al.(1993)]{con93} Condon, J. J., Helou, G., Sanders, D. B., \& Soifer, B. T. 1993, \aj, 105, 1730
\bibitem[Dale et al.(2006)]{dale06} Dale, D. A., et al. 2006, \apj, 646, 161
\bibitem[Dalgarno et al.(1999)]{dal99} Dalgarno, A.,Yan, M., \& Liu, W. 1999, \apjs, 125, 237
%\bibitem[de Messi\'{e}res et al.(2010)]{dem10} de Messi\'{e}res, G. E., O'Connell, R. W., McNamara, B. R., Donahue, M., Nulsen, P. E. J., Voit, G. M., \& Wise, M. W. 2010, ASPCS, 423, 105 %arXiv:0908.3445
\bibitem[Donahue et al.(2011)]{don11} Donahue, M., de Messi\'{e}res, G. E., O'Connell, R. W., Voit, G. M., Hoffer, A., McNamara, B. R., \& Nulsen, P. E. J. 2011, \apj, 732, 40
%\bibitem[Elmegreen \& Efremeov(1997)]{elm97} Elmegreen, B. G., \& Efremeov, Y. N. 1997, \apj, 480, 235
\bibitem[Egami et al.(2006)]{ega06} Egami, E., Rieke, G. H., Fadda, D., \& Hines, D. C. 2006, \apj, 652, L21
\bibitem[Fazio et al.(2004)]{faz04} Fazio, G., et al. 2004, \apjs, 154, 10 %IRAC
\bibitem[Flower \& Pineau des For\^{e}ts(2010)]{flower10} Flower, D. R., \& Pineau des For\^{e}ts, G. 2010, \mnras, 406, 1745
\bibitem[Gao et al.(2003)]{gao03} Gao, Y., Zhu, M., \& Seaquist, E. R. 2003, \aj, 126, 2171
\bibitem[Guillard et al.(2009)]{gui09} Guillard, P., Boulonger, F., Pineau des For\^{e}ts, G., \& Appleton, P. N. 2009, \aap, 502, 515
%\bibitem[Guillard et al.(2010a)]{gui10a} Guillard, P., Boulonger, F., Cluver, M. E., Appleton, P. N., Pineau des For\^{e}ts, G., \& Ogle, P., 2010a, \aap (in press)
\bibitem[Guillard et al.(2012)]{gui12} Guillard, P., et al. 2012, \apj (in press) %SQ CO paper 
\bibitem[Govoni \& Feretti(2004)]{gov04} Govoni, F., \& Feretti, L. 2004, Int. J. Mod. Phys. D, 13, 1549
%\bibitem[Haas et al.(2005)]{haa05} Haas, M., Chini, R., \& Klaas, U. 2005, \aap, 443, L17
\bibitem[Habart et al.(2011)]{hab11} Habart, E., et al. 2011, \aap, 527, A122
\bibitem[Habing(1968)]{hab68} Habing, H. J. 1968, \bain, 19, 421
\bibitem[Higdon et al.(2004)]{hig04} Higdon, S. J. U., et al. 2004, \pasp, 116, 975 %SMART
\bibitem[Higdon et al.(2006)]{hig06} Higdon, S. J. U., Armus, L., Higdon, J. L., Soifer, B. T. \& Spoon, H. W. W. 2006, \apj, 648, 323
\bibitem[Houck et al.(2004)]{hou04} Houck, J. R., et al. 2004, \apjs, 154, 18 %IRS
\bibitem[Helou et al.(2004)]{hel04} Helou, G., et al. 2004, \apjs, 154, 253
\bibitem[Jarrett et al.(1999)]{jar99} Jarrett, T. H., Helou, G., Van Buren, D., Valjavec, E., \& Condon, J. J. 1999, \aj, 118, 2132
%\bibitem[Kaufman et al.(2006)]{kau06} Kaufman,M. J., Wolfire, M. G., \& Hollenbach, D. J. 2006, \apj, 644, 283
\bibitem[Le Petit et al.(2006)]{lep06} Le Petit, F., Nehm\'{e}, C., Le Bourlot, J., \& Roueff, E. 2006, \apjs, 164, 506 %Meudon PDR code
\bibitem[Lisenfeld \& V\"{o}lk(2010)]{lis10} Lisenfeld, U., \& V\"{o}lk, H. 2010, \aap, 524, A27
\bibitem[Leroy et al.(2008)]{ler08} Leroy, A. K., Walter, F., Brinks, E., Bigiel, F., de Blok, W. J. G., Madore, B., \& Thornley, M. D. 2008, \aj, 136, 2782
\bibitem[Lutz et al.(2003)]{lutz03} Lutz, D., Sturn, E., Genzel, R., Spoon, H. W. W., Moorwood, A. F. M., Netzer, H., \& Sternberg, A. 2003, \aap, 409, 867
\bibitem[Mac Low(1999)]{mac99} Mac Low, M.-M. 1999, \apj, 524, 169
%\bibitem[Martin et al.(2005)]{mar05} Martin, D. C., et al. 2005, \apj, 619, L1 %GALEX
\bibitem[Nesvadba et al.(2010)]{nes10} Nesvadba, N. P. H., Boulanger, F., Salome, P., Guillard, P., Lehnert, M. D., Ogle, P., Appleton, P., Falgarone, E., \& Pineau des For\^{e}ts, G. 2010, \aap, 521, A65
%\bibitem[Morrissey et al.(2007)]{mor07} Morrissey, P., et al. 2007, \apjs, 173, 682 %GALEX calibration
\bibitem[Ogle et al.(2007)]{ogl07} Ogle, P., Antonucci, R., Appleton, P. N., \& Whysong, D. 2007, \apj, 657, 161
\bibitem[Ogle et al.(2010)]{ogl10} Ogle, P., et al. 2010, \apj, 724, 1193
\bibitem[Oka et al.(2005)]{oka05} Oka, T., Geballe, T. R., Gioto, M., Usuda, T., \& McCall, B. J. 2005, \apj, 632, 882
\bibitem[Rieke et al.(2004)]{rie04} Rieke, G. H., et al. 2004, \apjs, 154, 25 %MIPS
\bibitem[Rigopoulou et al.(2002)]{rig02} Rigopoulou, D., Kunze, D., Lutz, D., Genzel, R., \& Moorwood, A. F. M. 2002, \aap, 389, 374
\bibitem[Roman-Duval et al.(2010)]{rom10} Roman-Duval, J, Jackson, J. M., Heyer, M.; Rathborne, J., \& Simon, R. 2010, \apj, 723, 49
\bibitem[Roussel et al.(2007)]{rou07} Roussel, H., et al. 2007, \apj, 669, 959
\bibitem[S\'{a}nchez et al.(2010)]{san10} S\'{a}nchez, N., A\~{n}ez, N., Alfaro, E. J., \& Crone Odekon, M. 2010, \apj, 720, 541
\bibitem[Sanders et al.(2003)]{san03} Sanders, D. B., Mazzarella, J. M., Kim, D.-C., Surace, J. A., \& Soifer, B. T. 2003, \aj, 126, 1607
\bibitem[Sivanandam et al.(2009)]{siv09} Sivandandam, S., Rieke, M. J., Rieke, G. H. 2009, \apj, 717, 147
\bibitem[Smith \& Struck(2001)]{smi01} Smith, B. J., \& Struck, C. 2001, \aj, 121, 710
\bibitem[Smith et al.(2007a)]{smi07a} Smith, J. D. T., et al. 2007, \pasp, 119, 1133 %CUBISM
\bibitem[Smith et al.(2007b)]{smi07b} Smith, J. D. T., et al. 2007, \apj, 656, 770 %PAHFIT
\bibitem[Struck(1997)]{str97} Struck, C. 1997, \apjs, 113, 269
\bibitem[Tielens(2005)]{tie05} Tielens, A. G. G. M., 2005, The Physics and Chemistry of the Interstellar Medium (Cambridge, UK; Cambridge University Press)
\bibitem[Yusef-Zadeh et al.(2007)]{yus07}Yusef-Zadeh, F., Wardle, M., \& Roy, S. 2007, \apj, 665, L123
\bibitem[Werner et al.(2004)]{wer04} Werner, M. W., et al. 2004, \apjs, 154, 1 %Spitzer
\bibitem[Wilgenbus et al.(2000)]{wil00}Wilgenbus, D., Cabrit, S., Pineau des For\^{e}ts, G., \& Flower, D. R. 2000, \aap, 356, 1010 %o/p ratio at eq.
\bibitem[Zakamska(2010)]{zak10} Zakamska, N. L. 2010, \nat, 465, 60
\bibitem[Zhu et al.(2007)]{zhu07} Zhu, M., Gao, Y., Seaquist, E. R., \& Dunne, L. 2007, \aj, 134, 118
\end{thebibliography}
\end{document}